\nonstopmode
\documentclass[12pt]{article}

\usepackage{amsmath}
\usepackage{amsfonts}
\usepackage{amssymb}
\usepackage{bbm}
\usepackage{a4wide}
\usepackage{graphicx}
\graphicspath{{./}{../mp/}{../gpmp/}{../wtl/brute-force/}{../wtl/rhm/}}
\usepackage{hyperref}

\newcommand{\R}{\mathbb{R}}

\newcommand{\Z}{\mathbb{Z}}

\renewcommand{\P}{\mathbb{P}}
\newcommand{\dif}{\,{\rm d}}
\newcommand{\vol}{\mathop{\rm vol}}

\newcommand{\num}[1]{\mathcal{N}_{\text{#1}}}
\newcommand{\lww}{\Lambda_{\text{WW}}}

\usepackage{ifpdf}
\newcommand{\image}[2]{
  \ifpdf%
  \includegraphics[width=#1cm]{#2.pdf}%
  \else%
  \includegraphics[width=#1cm]{#2.eps}%
  \fi%
}

\title{%
  A geometric-probabilistic method for counting low-lying states in
  the Bousso-Polchinski Landscape%
}

\author{%
  C\'esar Asensio\thanks{\texttt{casencha}\textbf{@}\texttt{unizar.es}}%
  \qquad and \qquad %
  Antonio Segu\'{\i}\thanks{\texttt{segui}\textbf{@}\texttt{unizar.es}}%
  \\
  \\
  \emph{Theoretical Physics Department, University of Zaragoza}%
}

\begin{document}
\maketitle
\begin{abstract}
  We propose an accurate method for counting states of close to zero
  and positive cosmological constant in the Bousso-Polchinski
  Landscape.  This method is based on simple geometrical
  considerations on the high-dimensional lattice of quantized fluxes
  and on a probabilistic model (the ``random hyperplane'' model)
  which 
  provides a distribution of the values of the cosmological constant.
  Justification of the assumptions made in this model are offered by
  means of numerical experiments.
\end{abstract}

\section{Introduction}
\label{sec:intro}

One of the recent proposals to solve the cosmological constant problem
in cosmology is provided by string theory.  By dimensional reduction
from M-theory to 3+1 dimensions, vacua of the effective theory are
classified by means of a big number of quantized fluxes leading to an
enormous amount of metastable vacua, the Bousso-Polchinski (BP)
Landscape \cite{BP}.  The cosmological constant problem, namely the
smallness of the observed vacuum energy density in the universe
\cite{WW2,B-CC}, can be solved by the presence in this model of a huge
number of states of very small, positive cosmological constant,
together with a dynamical mechanism given by eternal inflation
\cite{EtInf} which allows to visit all the vacua.  An anthropic
selection is then advocated to explain the smallness of the observed
cosmological constant \cite{Anthr,WW}.

In order to quantify this selection a counting of accesible states in
the Landscape is needed.  Two ways of counting have been introduced so
far:
\begin{itemize}
\item The simplest Bousso-Polchinski count, which computes the volume
  of a spherical shell of small thickness in flux space and divides it
  by the volume of a cell.
\item The entropy count of Bousso-Yang, which computes the entropy of
  the occupation number of each flux assuming that they are
  independent.
\end{itemize}
In the following subsection we will briefly review these two counting
methods, and drawing on them we will propose an alternative one.

\subsection{The Bousso-Polchinski count}
\label{sec:BPcount}

The first estimate of this number is given by the BP count.  We will
now review this argument (see \cite{BP}).  A vacuum of the BP
Landscape is a node in a $J$-dimensional lattice ${\cal L}$ generated
by $J$ charges $q_1,\cdots,q_J$ determined by the sizes of the
three-cycles in the compactification manifold.  The lattice ${\cal L}$
is
\begin{equation}
  \label{eq:c1}
  {\cal L} = \bigl\{(n_1q_1,\cdots,n_Jq_J)\in\R^J\colon
  n_1,\cdots,n_J\in\Z\bigr\}\,.
\end{equation}
The $j$-th coordinate of a point in the lattice is an integer multiple
of the charge $q_j$, and therefore a vacuum is characterized by the
integer $J$-tuple $n=(n_1,\cdots,n_J)$.

A \emph{fundamental cell} (also called \emph{Voronoi
  cell}\footnote{Also called Wigner-Seitz cell in solid state physics,
  the Voronoi cell of a point $P$ in a discrete set $S$ of a metric
  space $M$ is the set of points of $M$ which are closer to $P$ than to
  any other point of $S$.}) $Q_n$ around a node $n$ in a lattice
${\cal L}$ is the subset of $\R^J$ which contains the points which are
closer to $n$ than to any other node of ${\cal L}$.  Thanks to the
discrete translational symmetry of our lattice (\ref{eq:c1}), all
fundamental cells in ${\cal L}$ are translates of the fundamental cell
around the origin $Q_O\equiv Q$, which we can parametrize in Cartesian
coordinates as a product of symmetric intervals
\begin{equation}
  \label{eq:c3}
  x\in Q \iff x = (x_1,\cdots,x_J) \text{ with }
  x_j\in\Bigl[-\frac{q_j}{2},\frac{q_j}{2}\Bigr]\,,
\end{equation}
i.e.
\begin{equation}
  \label{eq:c4}
  Q = \prod_{j=1}^J\Bigl[-\frac{q_j}{2},\frac{q_j}{2}\Bigr]\,.
\end{equation}
The cosmological constant of vacuum $n$ in the BP model is\footnote{We
  use reduced Planck units in which $8\pi G = \hbar = c = 1$.}
\begin{equation}
  \label{eq:c2}
  \Lambda(n) = \Lambda_0 + \frac{1}{2}\sum_{j=1}^J n_j^2q_j^2\,.
\end{equation}
In (\ref{eq:c2}), $\Lambda_0$ is an \emph{a priori} cosmological
constant or order $-1$.  Each value of $\Lambda>\Lambda_0$ defines a
spherical ball on the $J$-dimensional flux space of radius $R_\Lambda
= \sqrt{2(\Lambda-\Lambda_0)}$.  We call this ball ${\cal
  B}^{J}(\Lambda)$.  We take small values of the charges $q_j$
(natural values expected by BP are of order $\frac{1}{6}$) in such a
way that the ball can contain a huge number of fundamental cells.

The number of states in the Weinberg Window, that is the range of
values of the cosmological constant allowing the formation of
structures (like galaxies) needed for the formation of life as we know
it \cite{WW}, is obtained by computing
the volume of a thin spherical shell in flux space (the realization in
the BP Landscape of the Weinberg Window) divided by the volume of a
cell in the lattice:
\begin{equation}
  \label{eq:rh-3}
  \begin{split}
    \num{WW}
    &= \frac{\vol {\cal B}^J(\lww) - \vol {\cal B}^J(0)}{\vol Q}
    \approx \frac{1}{\vol Q}
    \left.
      \frac{\dif}{\dif\Lambda}\Bigl(\vol{\cal B}^J(\Lambda)\Bigr)
    \right|_{\Lambda=0}\lww \\
    &= \frac{1}{\vol Q}
    \left.
      \frac{\dif}{\dif\Lambda}
      \Bigl(\frac{R_\Lambda^J}{J}\vol{S}^{J-1}\Bigr)
    \right|_{\Lambda=0}\lww
    = \vol{S}^{J-1}\frac{R_0^{J-2}\lww}{\vol Q}\,,
  \end{split}
\end{equation}
where $R_0 = R_{\Lambda = 0} = \sqrt{2|\Lambda_0|}$, and the volume of
the $J-1$ dimensional sphere is
\begin{equation}
  \label{eq:c6}
  \vol S^{J-1} =
  \frac{2\pi^{\frac{J}{2}}}{\Gamma\bigl(\frac{J}{2}\bigr)}\,.
\end{equation}
This method can be naively expected to yield a good estimate when the
linear dimensions of the cell are small when compared to the thickness
of the shell; but this condition is not satisfied in the BP Landscape.
Nevertheless, the result of this counting formula is very good when
compared with actual numerical experiments.  We will derive the true
condition of validity of the BP count in our own framework in section
\ref{sec:N_S} below.

\subsection{The Bousso-Yang count}
\label{sec:BYcount}

The second estimate is given by the Monte Carlo numerical simulation by
Bousso-Yang (BY) in ref.~\cite{BY}.  They compute the number of states
by means of the Shannon entropy of the occupation number distribution
of each flux in a sample of states obtained in two ways:
\begin{itemize}
\item Drawing the states from a canonical ensemble distribution with
  spherical symmetry.
\item Drawing the states as the output of a decay chain using a dynamical
  relaxation mechanism.
\end{itemize}
This method has an advantage: it incorporates the dynamical relaxation
mechanism of Brown-Teitelboim (BT) \cite{BT-1,BT-2}, thus quantifying
the dynamical selection effect, but two drawbacks should be mentioned:
\begin{itemize}
\item It makes it necessary to choose particular values for the
  charges, thereby providing no explicit dependence of the computed
  number of states with the charges or the dimension, and
\item They assume that the states of low positive cosmological
  constant (referred to as \emph{penultimate} states) have fluxes
  which are considered independent random variables.  A hypothesis
  which they accept as not true but which they suggest how to correct.
\end{itemize}
We will briefly review their use of the Shannon entropy for counting.

For any subset $\Sigma$ having $\num{$\Sigma$}$ nodes in the Landscape
($\Sigma$ can be the set of penultimate states, for example), we can
define the uniform probability over $\Sigma$ as follows:
\begin{equation}
  \label{eq:rh-6}
  P_\Sigma(\widehat{n}=n) =
  \begin{cases}
    \text{const} & \text{if $n\in\Sigma$},\\
    0 & \text{if $n\notin\Sigma$},
  \end{cases}
\end{equation}
where $\widehat{n}$ is a random variable which can take values over
the whole Landscape $\mathcal{L}$ viewed as a subset of $\Z^J$, that
is, $\widehat{n}$ can be any integer $J$-tuple
$n=(n_1,\cdots,n_J)\in\mathcal{L}\subset\Z^J$ with equal probability,
namely $\frac{1}{\mathcal{N}_\Sigma}$.  The Shannon entropy $S_\Sigma$
of the uniform distribution $P_\Sigma$ satisfies
\begin{equation}
  \label{eq:rh-4}
  \num{$\Sigma$} = e^{S_\Sigma} = \exp\Bigl[
  -\sum_{n\in\Z^J}P_\Sigma(\widehat{n}=n)\log P_\Sigma(\widehat{n}=n)
  \Bigr]\,,
\end{equation}
as can be seen by subtituting (\ref{eq:rh-6}) into (\ref{eq:rh-4})
taking $\text{const} = \frac{1}{\mathcal{N}_\Sigma}$.  If $P_\Sigma$
were not constant over $\Sigma$, then it must have support larger than
$\Sigma$ in order to satisfy eq.~\eqref{eq:rh-4}.

If the fluxes were independent, this joint probability would split:
\begin{equation}
  \label{eq:rh-5}
  P_\Sigma(\widehat{n}=n) = \prod^J_{j=1}P_j(\widehat{n_j}=n_j)\,,
\end{equation}
and therefore the correspondent Shannon entropy would be simplified to
\begin{equation}
  \label{eq:rh-7}
S_\Sigma^{\text{indep}} =  -\sum_{j=1}^J \sum_{n\in\Z}
   P_j(\widehat{n_j}=n) \log  P_j(\widehat{n_j}=n)\,.
\end{equation}
Unlike $P_\Sigma$, the distributions $P_j(\widehat{n_j}=n)$ are much
simpler to estimate by sampling a small portion of $\Sigma$.  But, as
long as they are not constant, its support covers a much larger region
than $\Sigma$, and furthermore this region has the symmetry of the
cell $Q$ rather than the symmetry of $\Sigma$, so we can expect the
simplified entropy $S_\Sigma^{\text{indep}}$ of eq.~\eqref{eq:rh-7} to
be \emph{much} larger than the true entropy $S_\Sigma$.

A numerical experiment can be illustrative.  Needless to say, the only
way to compute the correct uniform probability $P_\Sigma$ is to
exhaustively compile all the elements of the set $\Sigma$, and this is
\emph{not} possible over a realistic Landscape.  So we have taken a
very simple model of $J=3$ fluxes with $\Lambda_0=-1$ and charges $q_1
= 0.02988$, $q_2=0.04988$, $q_3=0.06988$, and as the set $\Sigma$ we
have considered the \emph{secant} states (see next subsection).  A
brute-force count of this set is easily carried out by enumerating all
states in it, resulting in $S_\Sigma = 9.79222$ or
\begin{equation}
  \label{eq:rh-9}
  \num{$\Sigma$} = 17,894\quad\text{states}.
\end{equation}
The same enumeration of all states allows us to compute the three
probability distributions $P_j(\widehat{n_j}=n_j)$, and from these we
obtain the additive entropy $S_\Sigma^{\text{indep}} = 12.27932$, and
a state count of
\begin{equation}
  \label{eq:rh-10}
  \num{$\Sigma$}^{\text{indep}} = e^{S_\Sigma^{\text{indep}}} = 215,199.1\,,
\end{equation}
which, indeed, means an enormous difference.

It must be stressed that the goal of the BY work is to quantify the
dynamical selection effect of the BT decay chain, and they do this by
computing the quotient of the number of states with and without
dynamics.  While both numbers are affected by the independency
error, their quotient may be free of errors. We are interested in
studying this point in our future research.

\subsection{Our counting method}
\label{sec:our-method}

\begin{figure}[htbp]
  \centering
  \includegraphics[width=8cm]{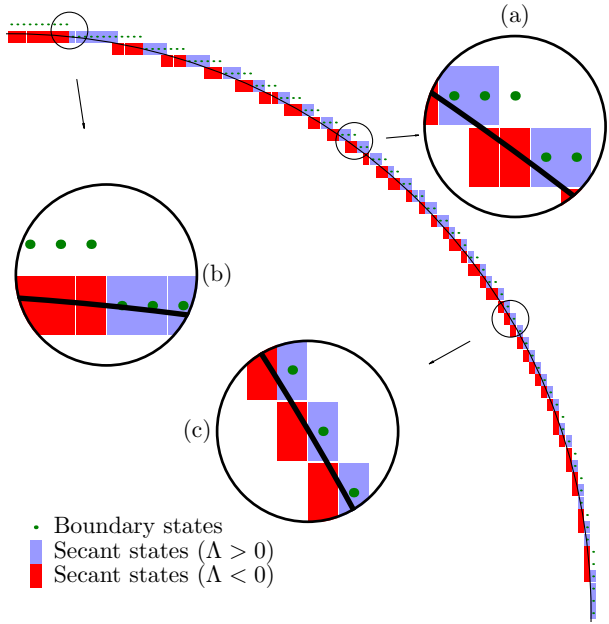}
  \caption{States are shown on the $\Lambda=0$ surface in a $J=2$ BP
    Landscape.  We see secant states which are not boundary and
    viceversa (a), regions in which S:B are in a 1:1 ratio (b) and
    regions in which S:B are in a 2:1 ratio (c).}
\label{fig:b-s}
\end{figure}

Our proposal is based on the following kinds of states one may
encounter near the null cosmological constant surface in flux space:
\begin{itemize}
\item \textbf{Boundary} (or \textbf{penultimate} after BY) are
  those states in which a BT decay chain can end before jumping into
  the negative cosmological constant sea.  So we define a boundary
  state as one having
  \begin{itemize}
  \item [(1)] positive cosmological constant, and
  \item [(2)] at least one neighbor of negative cosmological constant.
  \end{itemize}
\item \textbf{Secant} states have the property that their Voronoi cell
  in flux space has non-empty intersection with the null cosmological
  constant surface in flux space.  Note that a secant state may have
  negative cosmological constant.
\end{itemize}
These two categories are not equivalent; a boundary state may not be
secant if it is far enough from the null cosmological constant
surface, and a secant state may not be boundary if it has negative
cosmological constant.  So we are interested mainly in the states which
are both secant \textbf{and} boundary, because all the states in the
Weinberg Window are in this category.  Figure \ref{fig:b-s}
illustrates the differences between these two kinds of cell.

Our strategy would be as follows.  We will count the states in the
Weinberg Window using the following elementary formula:
\begin{equation}
  \label{eq:rh-1}
  \num{WW}
  = \frac{1}{2}\num{S} P(\Lambda\in\text{WW})\,,
\end{equation}
where
\begin{itemize}
\item $\num{WW}$ is the number of states in the Weinberg Window,
\item $\num{S}$ is the total number of secant states,
\item the $\frac{1}{2}$ factor is the (first-order) approximate
  fraction of positive cosmological constant secant states,
\item $P(\Lambda\in\text{WW})$ is the probability that a random secant
  state has a positive cosmological constant in the Weinberg Window.
  If we call a number of the size of $10^{-120}$ $\lww$, then
  \begin{equation}
    \label{eq:rh-2}
    P(\Lambda\in\text{WW}) = P(0 < \Lambda < \lww)\,,
  \end{equation}
  where the probability must be computed using the distribution of the
  cosmological constant as a random variable over all the secant
  states.
\end{itemize}
In section \ref{sec:N_S} we first compute the number $\num{S}$ and
check it with simple models with two or three fluxes, where
brute-force counting is feasible.  In the following section we propose
a probabilistic model (and check it against numerical data) which leads
to the distribution of the values of the cosmological constant
restricted to the secant states (section \ref{sec:rhm}).  We use this
distribution to compute the probability (\ref{eq:rh-2}) in section
\ref{sec:N_WW}.  Surprisingly, our results show that this more precise
counting method yields the same result as the BP count.  Finally, we
summarize the conclusions in section \ref{sec:conc}.

\section{Counting secant states}
\label{sec:N_S}

\begin{figure}
  \centering
  \image{8}{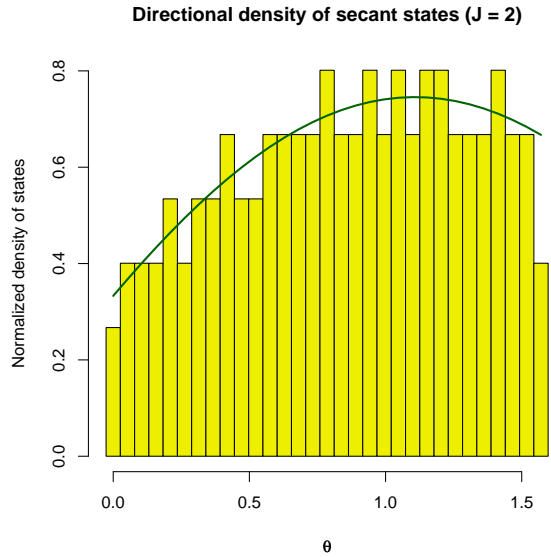}
  \caption{Directional density of states in the BP Landscape shown in
    figure \ref{fig:b-s} computed numerically compared to the
    continuous version $\nu(\theta)$ of equation \eqref{eq:c0-2}.  All
    plots in this paper regarding statistical analysis were done using
    R \cite{R-proj}.}
\label{fig:bf-hist-1}
\end{figure}

We start with the observation that the number of secant states which we
can find by looking from the origin in a given direction is not a
constant, as can be seen in the histogram shown in figure
\ref{fig:bf-hist-1}.
In this figure, we can also see the theoretical density of states which
we now derive for the $J=2$ case.

\begin{figure}[htbp]
  \centering
  \includegraphics[width=8cm]{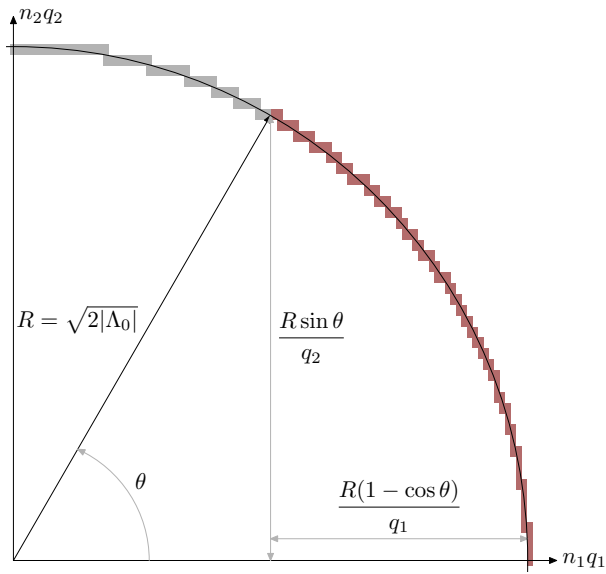}
  \caption{Construction of $N(\theta)$, equation \eqref{eq:c0}.}
\label{fig:bf-fig-2}
\end{figure}

Let $N(\theta)$ be the number of secant states on the first quadrant
of a BP Landscape which are between the 1-axis and a straight line
drawn at an angle $\theta$, as in figure \ref{fig:bf-fig-2}.
The number of states along the arc between the 1-axis and the
$\theta$-angle line can be accurately approximated by the length of
the segments which the states along the $\Lambda=0$ circle are
covering.  These lengths can be straightened as shown in figure
\ref{fig:bf-fig-2}, to yield (here we call $R=R_0=\sqrt{2|\Lambda_0|}$
for convenience)
\begin{equation}
  \label{eq:c0}
  N(\theta) = \frac{R\sin\theta}{q_2} + \frac{R(1-\cos\theta)}{q_1}\,.
\end{equation}
Its derivative is the directional density of states:
\begin{equation}
  \label{eq:c0-2}
  \nu(\theta) = \frac{\dif N}{\dif\theta} =
  R\Bigl(\frac{\cos\theta}{q_2} + \frac{\sin\theta}{q_1}\Bigr) =
  \frac{R}{\vol Q}\bigl(q_1\cos\theta + q_2\sin\theta\bigr) = \frac{R
    q\cdot\upsilon}{\vol Q}\,.
\end{equation}
In equation \eqref{eq:c0-2}, $\vol Q = q_1q_2$,
$\upsilon=(\cos\theta,\sin\theta)$ and $q=(q_1,q_2)$.  Only the first
quadrant is considered here and the absolute value on the components
of $\upsilon$ must be taken if we want to extend equation
\eqref{eq:c0-2} to all quadrants.  The formula thus obtained is
plotted in figure \ref{fig:bf-hist-1} over the numerical data, giving
an accurately approximated density.  Of course we can estimate the
total number of secant states by
\begin{equation}
  \label{eq:c0-3}
  \mathcal{N}^{\text{theo}}_S = 4\int^{\frac{\pi}{2}}_0
  \nu(\theta)\dif\theta = 
  \frac{4R(q_1+q_2)}{q_1q_2}\,.
\end{equation}
Using the $J=2$ values $q_1 = 0.01494$, $q_2=0.02994$ and $\Lambda_0 =
-1$, we have $N^{\text{theo}}_S = 4\times 141.8945 = 567.5779$, being
the correct number, obtained by an exhaustive search, of 143 states,
2 of them on the axes, which gives a total of
\begin{equation}
  \label{eq:c0-4}
  \mathcal{N}^{\text{num}}_S = 4\times 141 + 4 = 568 \approx 567.5779
  =  \mathcal{N}^{\text{theo}}_S\,, 
\end{equation}
The two results clearly agree.

For a model with $J=3$, we can see that the Voronoi cells of the
secant states on the first octant project their faces over the first
quadrants of planes 1-2, 1-3 and 2-3.  The area of these quadrants is
$\frac{\pi}{4}R^2$ and the number of states needed to cover this total
area is
\begin{equation}
  \label{eq:c0-5}
  \mathcal{N}^{\text{theo}}_S = \frac{\pi R^2}{4q_1q_2} +
  \frac{\pi R^2}{4q_1q_3} +
  \frac{\pi R^2}{4q_2q_3} = \frac{\pi R^2}{4\vol
    Q}\bigl(q_1+q_2+q_3\bigr)\,.
\end{equation}
Let us now take the density \eqref{eq:c0-2} and extend it to $J=3$.
The correct scaling will be
\begin{equation}
  \label{eq:c0-6}
  \nu(\upsilon) = \frac{R^2}{\vol Q}q\cdot\upsilon\,,
\end{equation}
where $\upsilon$ is a norm-one vector on the $S^2$ sphere.  Let us
integrate this function on the first octant:
\begin{equation}
  \label{eq:c0-7}
  \frac{1}{8}\int_{S^2}\nu(\upsilon)\dif\Omega_2(\upsilon) =
  \frac{R^2}{q_1q_2q_3}
  \int^{\frac{\pi}{2}}_0\sin\theta\dif\theta\int^{\frac{\pi}{2}}_0\dif\phi
  \bigl(q_1\sin\theta\cos\phi + q_2\sin\theta\sin\phi + q_3\cos\theta\bigr)\,.
\end{equation}
The $\phi$-integrals have the values 1, 1, $\frac{\pi}{2}$, and the
$\theta$-integrals have the values $\frac{\pi}{4}$, $\frac{\pi}{4}$,
$\frac{1}{2}$, so that all integrals coincide and the result is
\eqref{eq:c0-5}.

Using the values $q_1=0.01494$, $q_2=0.02494$, $q_3=0.03494$,
$\Lambda_0=-1$ for the $J=3$ model, we have an 8-octant total number
of secant states of
\begin{equation}
  \label{eq:c0-8}
  \mathcal{N}^{\text{theo}}_S = \frac{8\pi R^2}{4\vol Q}\bigl(q_1 +
  q_2 + q_3\bigr) = \frac{4\pi\times 0.07482}{1.301877\cdot10^{-5}} =
  72220.02\,.
\end{equation}
A brute-force search of the secant states gives 9205 states on the
first octant.  But 3 are on the axes, $96+135+152=383$ on the
coordinate planes and the rest ($9205 - 383 - 6 = 8819$) outside, so
the total number of secant states found in this model is
\begin{equation}
  \label{eq:c0-9}
  \mathcal{N}^{\text{num}}_S = 2\times 3 + 4\times 383 + 8\times 8819
  = 72090\,.
\end{equation}
The agreement in this case is not complete but it is satisfactory.

The discrepancy may be caused for the sampling method of the secant
states in the brute-force search.  It consists on uniformly sampling
the $S^{2}_0$ sphere of radius $\sqrt{2|\Lambda_0|}$ correspondent to
$\Lambda=0$; each point thus sampled hits on the intersection between
the sphere $S^2_0$ and the Voronoi cell $Q_n$ of certain secant state
$n$.  Therefore, the probability for this secant state $n$ to be
selected by this method is proportional to the \emph{area}
($(J-1)$-volume) of the intersection:
\begin{equation}
  \label{eq:c0-10}
  P(n) = \frac{\vol(Q_n\cap S^2_0)}{\vol S^2_0}\,.
\end{equation}
If we call the set of secant states $\mathcal{S}$, the complete and
disjoint partition (tessellation) which the Voronoi cells induce on the
sphere guarantees that this probability is normalized:
\begin{equation}
  \label{eq:c0-11}
  \sum_{n\in\mathcal{S}}P(n) = 1\,.
\end{equation}
Thus, only the states with bigger intersection area will be selected
in a sample taken by this method.  In addition, the intersection
volume is positive but it does not have a positive lower bound, so in
principle there can be arbitrarily small intersections which will
\emph{not} be detected by this method.  In the $J=2$ case, $10^4$
points were needed to find the existing 143 secant states (see figure
\ref{fig:b-s}), but using $10^8$ points on the $J=3$ case yielded 9134
states of the 9205 revealed in the $10^9$ sample.  If the set
$\mathcal{S}$ contains states with a probability below $10^{-9}$,
these will \textbf{not} be found using this method.  So the agreement
may increase by taking a sample size greater than $10^9$.

Let us extend the directional density of states \eqref{eq:c0-6} to an
arbitrary dimension $J$ and to all ``quadrants'':
\begin{equation}
  \label{eq:c0-12}
  \nu(\upsilon) = \frac{R^{J-1}}{\vol Q}\,q\cdot|\upsilon|\,,
\end{equation}
where $\upsilon=(\upsilon_1,\cdots,\upsilon_J)\in S^{J-1}$,
$|\upsilon|=(|\upsilon_1|,\cdots,|\upsilon_J|)$ and
$q=(q_1,\cdots,q_J)$.  The number of secant states is
\begin{equation}
  \label{eq:c0-13}
  \mathcal{N}_\mathcal{S} =
  \int_{S^{J-1}}\nu(\upsilon)\dif\Omega_{J-1}(\upsilon) = 
  \frac{R^{J-1}}{\vol Q}\sum^J_{i=1}q_i \int_{S^{J-1}}
  |\upsilon_i|\dif\Omega_{J-1}(\upsilon)\,. 
\end{equation}
The integration measure on the sphere $\dif\Omega_{J-1}(\upsilon)$ is
invariant under $SO(J)$ rotations, and there is a rotation which can
transform a given coordinate $\upsilon_i$ in the simplest of them,
$\upsilon_J$, in the following choice of coordinates for the sphere:
\begin{equation}
  \label{eq:c0-14}
  \upsilon = w\sin\theta + e_J\cos\theta\,, \quad
  w\in S^{J-2}\,,\quad
  \theta\in[0,\pi]\,,
\end{equation}
where $e_J$ is the last vector in a $\R^J$ basis.  The integration
measure reduces to
\begin{equation}
  \label{eq:c0-15}
  \dif\Omega_{J-1}(\upsilon) =
  \sin^{J-2}\theta\dif\theta\dif\Omega_{J-2}(w)\,, 
\end{equation}
so that
\begin{equation}
  \label{eq:c0-16}
  \int_{S^{J-1}}
  |\upsilon_i|\dif\Omega_{J-1}(\upsilon) =
  2\int^{\frac{\pi}{2}}_0\cos\theta\sin^{J-2}\theta\dif\theta
  \int_{S^{J-2}}\dif\Omega_{J-2}(w)
  = \frac{2\vol S^{J-2}}{J-1}\,.
\end{equation}
We have then
\begin{equation}
  \label{eq:c0-17}
  \mathcal{N}_\mathcal{S}
  = \frac{2 R^{J-1}\vol S^{J-2}}{(J-1)\vol Q}\sum^J_{i=1}q_i 
  = \frac{2 R^{J-1}\overline{q} \vol S^{J-2}}{\vol Q}\,. 
\end{equation}
In a somewhat nonstandard way, we have defined the quantity
\begin{equation}
  \label{eq:c0-19}
  \frac{1}{J-1}\sum^J_{i=1}q_i = \overline{q}\,.
\end{equation}
Equations \eqref{eq:c0-8} and \eqref{eq:c0-3} are special cases of
\eqref{eq:c0-17}.  A more explicit form of \eqref{eq:c0-17}, using
$R=\sqrt{2|\Lambda_0|}$ and $\vol S^{J-2} =
2\pi^{\frac{J-1}{2}}/\Gamma(\frac{J-1}{2})$ is
\begin{equation}
  \label{eq:c0-18}
  \mathcal{N}_\mathcal{S} = 
  2\,
  \frac{[2\pi|\Lambda_0|]^{\frac{J-1}{2}}}{\Gamma\bigl(\frac{J+1}{2}\bigr)}
  \,
  \frac{\sum^J_{i=1}q_i}
  {\prod^J_{i=1}q_i}
  \,.
\end{equation}
Let us now derive the condition of validity of formulae
(\ref{eq:c0-17},~\ref{eq:c0-18}).  Note that we are summing the
number of times that a fundamental cell fits in a $(J-1)$-quadrant; in
order for this number to represent the actual number of secant states,
the excess $(J-1)$-volume of the cells trying to fit the boundary must
be very small when compared with the $(J-1)$-volume of the
$(J-1)$-quadrant.  The $(J-1)$-volume of the cell projection is
different in each direction, so we can compute a kind of mean value of
the projection volume as
\begin{equation}
  \label{eq:c0-19-1}
  \mu^{J-1} = \frac{1}{J}\sum^J_{j=1}\prod^J_{\substack{i=1\\i\ne
      j}}q_i = \frac{\widehat{q}^J}{q_H}\,,
\end{equation}
where $\widehat{q}$ and $q_H$ are respectively the geometric and
harmonic means of the charges.  This mean projection volume must be
much less than the $(J-1)$-volume of a single $(J-1)$-quadrant, that
is,
\begin{equation}
  \label{eq:c0-19-2}
  \mu^{J-1} \ll
  \frac{1}{2^{J-1}}\,
  \frac{R^{J-1}}{J-1}\,\vol S^{J-2}\,.
\end{equation}
After substituting $R=\sqrt{2|\Lambda_0|}$ and $\vol S^{J-2} =
2\pi^{\frac{J-1}{2}}/\Gamma(\frac{J-1}{2})$, we
find\footnote{Incidentally, this relation resembles the so-called
  t'Hooft limit in field theory, in which the number $N$
  characterizing the gauge group tends to infinity and the Yang-Mills
  coupling constant $g_{\text{YM}}$ vanishes with the product
  $Ng_{\text{YM}}^2$ (the t'Hooft coupling) held fixed.}
\begin{equation}
  \label{eq:c0-19-3}
  J\frac{\mu^2}{|\Lambda_0|} \ll \pi e \approx 8.539734\,.
\end{equation}

\section{The random hyperplane model}
\label{sec:rhm}

We assume two basic features of the secant states in a BP Landscape:
\begin{itemize}
\item The Voronoi cells of the secant states are small enough to
  replace the $\Lambda=0$ sphere which intersects the cell by its
  tangent hyperplane.
\item The orientations of the hyperplane intersecting the cell are
  random if one picks a secant state at random.
\end{itemize}
In this way, we propose to study the set of secant states by choosing
a probability measure on the secant hyperplane set.  In this section
we first parametrize the hyperplane space, and then we choose a
probability measure on it.

\subsection{The hyperplane space}
\label{sec:hyp-space}

We define $H$ as the set of all hyperplanes in $\R^J$ and $H_Q$ as the
set of all hyperplanes with non-empty intersection with $Q$:
\begin{equation}
  \label{eq:c12}
  H_Q = \bigl\{h\in H\colon Q\cap h\ne\varnothing\bigr\}\,.
\end{equation}
In order to specify a hyperplane $h\in H_Q$ we must choose, first of
all, a normal vector of unit norm, $\upsilon\in S^{J-1}$.  A point in
$h$ must also be given; but there is an infinity of possible choices
here, and we must provide a unique prescription, for example the point
$z\in h$ closest to the origin $O$ (chosen here as the center of $Q$).
Note that this point $z$ may lie inside or outside $Q$.

The pair $(z,\upsilon)$ has a lot of redundant information, because
the directions of the vector $Oz=(z_1,\cdots,z_J)$ and
$\upsilon=(\upsilon_1,\cdots,\upsilon_J)$ coincide, that is,
$Oz=\pm|Oz|\upsilon$.  Therefore, we must retain only the norm of $Oz$
and the whole $\upsilon$; letting $\rho=|Oz|$, we can identify $h$
with the pair $(\rho, \upsilon)$.  Also, we must take into account the
fact that $\upsilon$ and $-\upsilon$ represent different hyperplanes
if $\rho\ne0$, but the same hyperplane if $\rho=0$, so the vector
$\upsilon$ is defined up to a sign in the case $\rho=0$.

An alternative equivalent prescription is to consider directions up to
a sign for all $\rho$ (the point $\upsilon$ is on $S^{J-1}$ with
antipodal points identified, i.e.~a point on the projective space
$\P^{J-1}$).  In this case, $\rho$ represents not only the minimum
distance from the hyperplane to the origin $O$, but also the position
of the hyperplane ``above'' or ``below'' the origin.  We find the
former point of view more adequate.

The domain of definition of $\rho$ is a direction-dependent positive
interval $I_\upsilon = [0,\sigma(\upsilon)]$, and is determined by the
restriction that the hyperplane it represents has non-empty
intersection with $Q$.  In particular, the point $z$ of minimum
distance can lie outside $Q$; thus, the ``hyperplane space'', when
compared with ``physical'' space, comprises a larger region than $Q$.

\begin{figure}[htbp]
  \centering
  \includegraphics[scale=1]{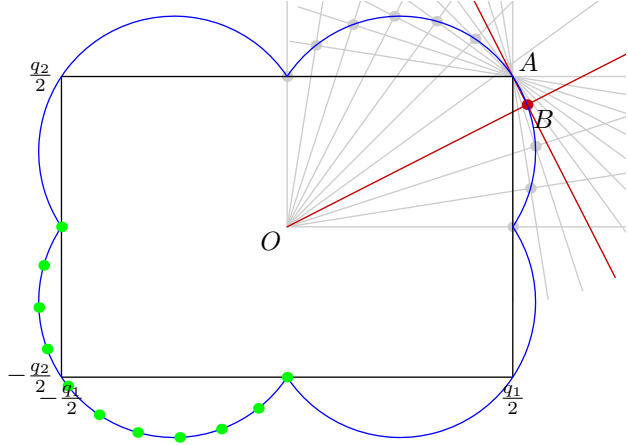}
  \caption{Construction of the hyperplane space in the $J=2$ case.
    The blue contour comprises all points which represent a different
    hyperplane intersecting the cell.  Here, the corner $A$ is used to
    construct the first quadrant part of the contour, and $\sigma =
    |OB|$.  The green points have been generated using formula
    \eqref{eq:c63} on the third quadrant.}
\label{fig:H_Q}
\end{figure}

\begin{figure}[htbp]
  \centering
  \includegraphics[width=10cm]{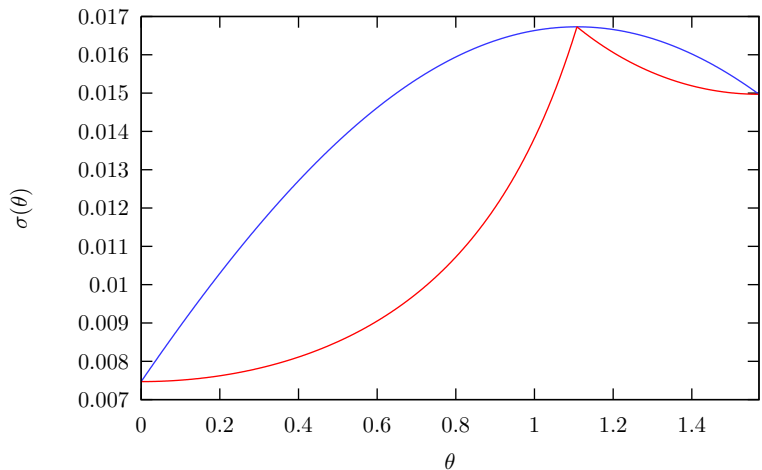}
  \caption{The function $\sigma$ in the $J=2$ case, with $q_1 =
    0.01494$, $q_2=0.02994$ and $\upsilon=(\cos\theta,\sin\theta)$.
    The cell boundary is also displayed.  The corner is at
    $\theta_0=\arctan\frac{q_2}{q_1}$, and is the maximum of
    $\sigma$.}
\label{fig:sigma}
\end{figure}

We now compute $\sigma(\upsilon)$, defined as the maximum distance to
the origin of the closest point to the origin of a hyperplane with
non-empty intersection with the cell $Q$.  Clearly, the corners of the
cell are the most far away points in $Q$, so for each direction
$\upsilon$, the last (most distant) hyperplane orthogonal to
$\upsilon$ and with non-empty intersection with $Q$ must contain one
of the $2^J$ corners of the cell.  The equation of such a hyperplane
is
\begin{equation}
  \label{eq:c62}
  \upsilon\cdot(x - c_\upsilon) = 0\,,
\end{equation}
where $c_\upsilon=\frac{1}{2}(s_1q_1,\cdots,s_Jq_J)$ is the unique
corner out of $2^J$ which belongs to the same $J$-quadrant as
$\upsilon$, and the $s_j$ are signs $\pm1$, indeed the same signs of
the components of $\upsilon$, that is,
$\upsilon=(s_1|\upsilon_1|,\cdots,s_J|\upsilon_J|)$.  This hyperplane
is already in normal form, so its minimum distance to the origin is
\begin{equation}
  \label{eq:c63}
  \sigma(\upsilon) = \upsilon\cdot c_\upsilon
  = \frac{1}{2}\sum^J_{j=1} q_j s_j^2 |\upsilon_j| =
  \frac{1}{2}\sum^J_{j=1}q_j|\upsilon_j| = \frac{1}{2}q\cdot|\upsilon|\,.
\end{equation}
Given a direction $\upsilon$, the closest point can be found as the
intersection of the line $x=\upsilon t$ with the hyperplane
\eqref{eq:c62}.  This method is used to construct the hyperplane space
for the $J=2$ case in figure \ref{fig:H_Q}, where it is compared with
formula \eqref{eq:c63}.
The function $\sigma(\upsilon)$ can be plotted for $J=2$ and $q_1 =
0.01494$, $q_2=0.02994$, versus a polar angle $\theta$ defined as
$\upsilon=(\cos\theta,\sin\theta)$, see figure \ref{fig:sigma}.

\subsection{A probability measure in the hyperplane space}
\label{sec:unif-prob-hyp-space}

Now we need to define a probability measure on the hyperplane space
just parametrized.  The simplest choice is the uniform probability on
$H_Q$:
\begin{equation}
  \label{eq:rh-8}
  \dif P(h) = \dif P(\rho,\upsilon)
  = K^{-1}\chi_{H_Q}(h)\dif\rho\dif\Omega_{J-1}(\upsilon)\,,
\end{equation}
where $\chi_{H_Q}(h)$ is the characteristic function on $H_Q$, whose
effect is simply to restrict the integration domain to $H_Q$, and
$\dif\Omega_{J-1}(\upsilon)$ is the volume measure in $S^{J-1}$.
$K$ is a normalization constant:
\begin{equation}
  \label{eq:c19}
  K = \int_{H_Q}\dif \rho \dif\Omega_{J-1}(\upsilon)\,.
\end{equation}
We can compute $K$ by first integrating out the $\rho$ variable in
\eqref{eq:c19}:
\begin{equation}
  \label{eq:64}
  K = \int_{S^{J-1}}
  \int^{\sigma(\upsilon)}_0\dif\rho\dif\Omega_{J-1}(\upsilon)
  = \int_{S^{J-1}} \sigma(\upsilon)\dif\Omega_{J-1}(\upsilon)
  \,.
\end{equation}
By comparing equations \eqref{eq:c0-12} and \eqref{eq:c63} we observe
the following relation between the maximum distance of a hyperplane in
a given direction $\sigma(\upsilon)$ and the directional density of
states $\nu(\upsilon)$:
\begin{equation}
  \label{eq:rh-11}
  \sigma(\upsilon) = \frac{1}{2}\frac{\vol Q}{R^{J-1}}\ \nu(\upsilon)\,,
\end{equation}
so that using the definition of the density of states
\begin{equation}
  \label{eq:rh-12}
  \mathcal{N}_\mathcal{S} =
  \int_{S^{J-1}}\nu(\upsilon)\dif\Omega_{J-1}(\upsilon) 
\end{equation}
we have
\begin{equation}
  \label{eq:rh-13}
  K = \int_{S^{J-1}}\sigma(\upsilon)\dif\Omega_{J-1}(\upsilon) =
  \frac{\mathcal{N}_\mathcal{S} \vol Q}{2\,R^{J-1}}
  = \overline{q}\,\vol S^{J-2}
  \,.
\end{equation}
In the last formula we have substituted the expression for
$\mathcal{N}_\mathcal{S}$ found in equation~\eqref{eq:c0-17}.

The main reason for choosing the uniform probability measure is its
simplicity.  The physical reason is that among all measures in a given
compact space, the uniform one has maximum Shannon entropy.  In the
following, we will justify our choice by more quantitative means.

First of all, the marginal probability in the $\upsilon$ variable
associated to the uniform measure is proportional to the directional
density of states $\nu(\upsilon)$, with a normalization-to-one factor:
\begin{equation}
  \label{eq:rh-14}
  \int_{\rho\in[0,\sigma(\upsilon)]}\dif P(\rho,\upsilon) =
  K^{-1}\dif\Omega_{J-1}(\upsilon)\int_0^{\sigma(\upsilon)} \dif\rho =
  \frac{2\,R^{J-1}}{\mathcal{N}_\mathcal{S} \vol Q} \,\sigma(\upsilon)
  \dif\Omega_{J-1}(\upsilon)
  =
  \frac{\nu(\upsilon)}{\mathcal{N}_\mathcal{S}}
  \dif\Omega_{J-1}(\upsilon)
  \,.
\end{equation}
Therefore, the uniform probability reproduces the correct directional
density of states as observed in the numerical experiments.

\begin{figure}
  \centering
  \image{7}{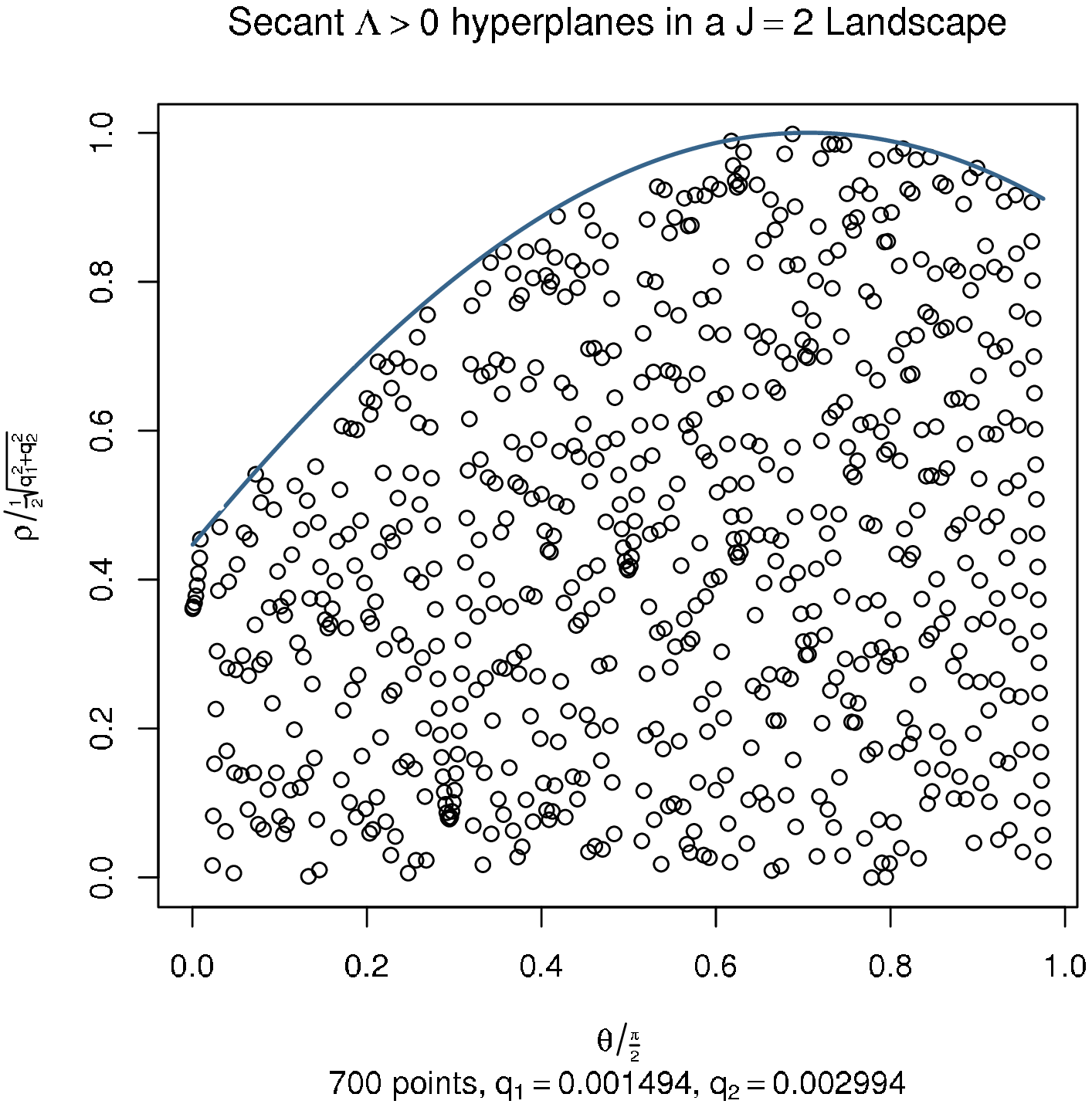}\qquad%
  \image{7}{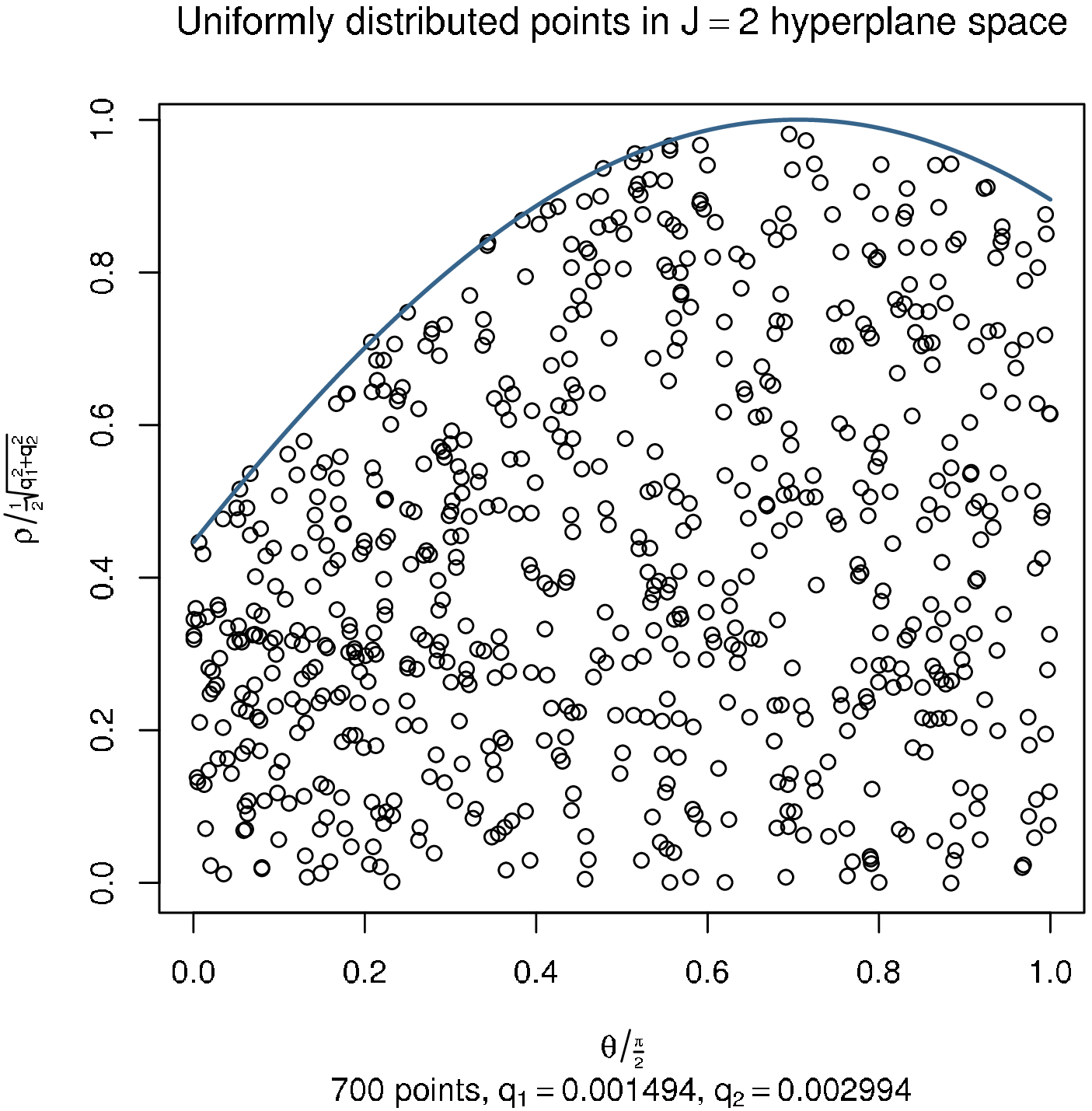}%
  \caption{Actual secant hyperplane plot of a $J=2$ Landscape in
    hyperplane space (\emph{geometric} sample, left) and a uniformly
    distributed sample in the same domain (\emph{simulated} sample,
    right).  Note the structures in the former, and the bigger voids
    and clusters in the latter.}
  \label{fig:sec-vs-sim}
\end{figure}

On the other hand, we can plot instances of the set of secant
hyperplanes and compare them with simulated uniform points in
hyperplane space.  The simplest of such plots is for $J=2$; with
$\Lambda_0=-1$ and charges $q_1=0.001494$, $q_2=0.002994$, we show the
$(\theta,\rho)$ points from the actual secant hyperplanes versus a
simulated set of uniform points of equal sample size in figure
\ref{fig:sec-vs-sim}.

The geometrical nature of the secant hyperplane sample is revealed in
the structures shown in fig.~\ref{fig:sec-vs-sim} (left), which
introduces correlations in the spatial sequence.  In contrast, a
uniform sample in hyperplane space (fig.~\ref{fig:sec-vs-sim}, right)
shows no correlations, and the only structures which we can see are
voids and clusters bigger than the ones present in the former case.

Both samples thus obtained (the secant hyperplane set, or
\emph{geometric} sample, and the uniformly distributed points, or
\emph{simulated} sample) cover the hyperplane space, so we can
approximate the distribution of the former set of points by a uniform
probability, thus neglecting the spatial correlations.

\begin{figure}
  \centering
  \image{8}{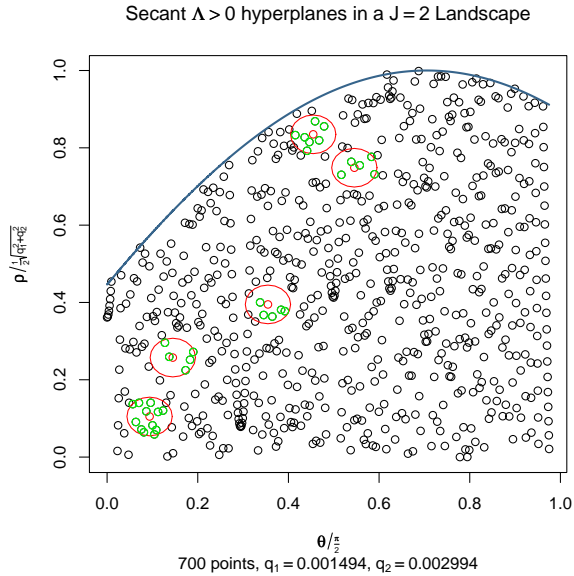}
  \caption{Illustration of the circle method to measure the voids and
    clusters distribution.  Small circles are thrown randomly inside
    hyperplane space.  Green points are the sample points inside the
    circles; red points are the random centers of the circles.}
  \label{fig:circle-method}
\end{figure}

But the efficiency of the covering is not equal in both cases: the
geometrical sample shows smaller voids and clusters than the simulated
one.  This can be seen by counting the number of points inside a
circle of random position and given radius inside the hyperplane space
in both cases (fig.~\ref{fig:circle-method} illustrates this ``circle
method'').
When this is done for a big number of such circles, the voids and
clusters induce a fluctuation in the number of ``inner'' points, which
will be greater if the inhomogeneities are bigger.
\begin{figure}
  \centering
  \image{7}{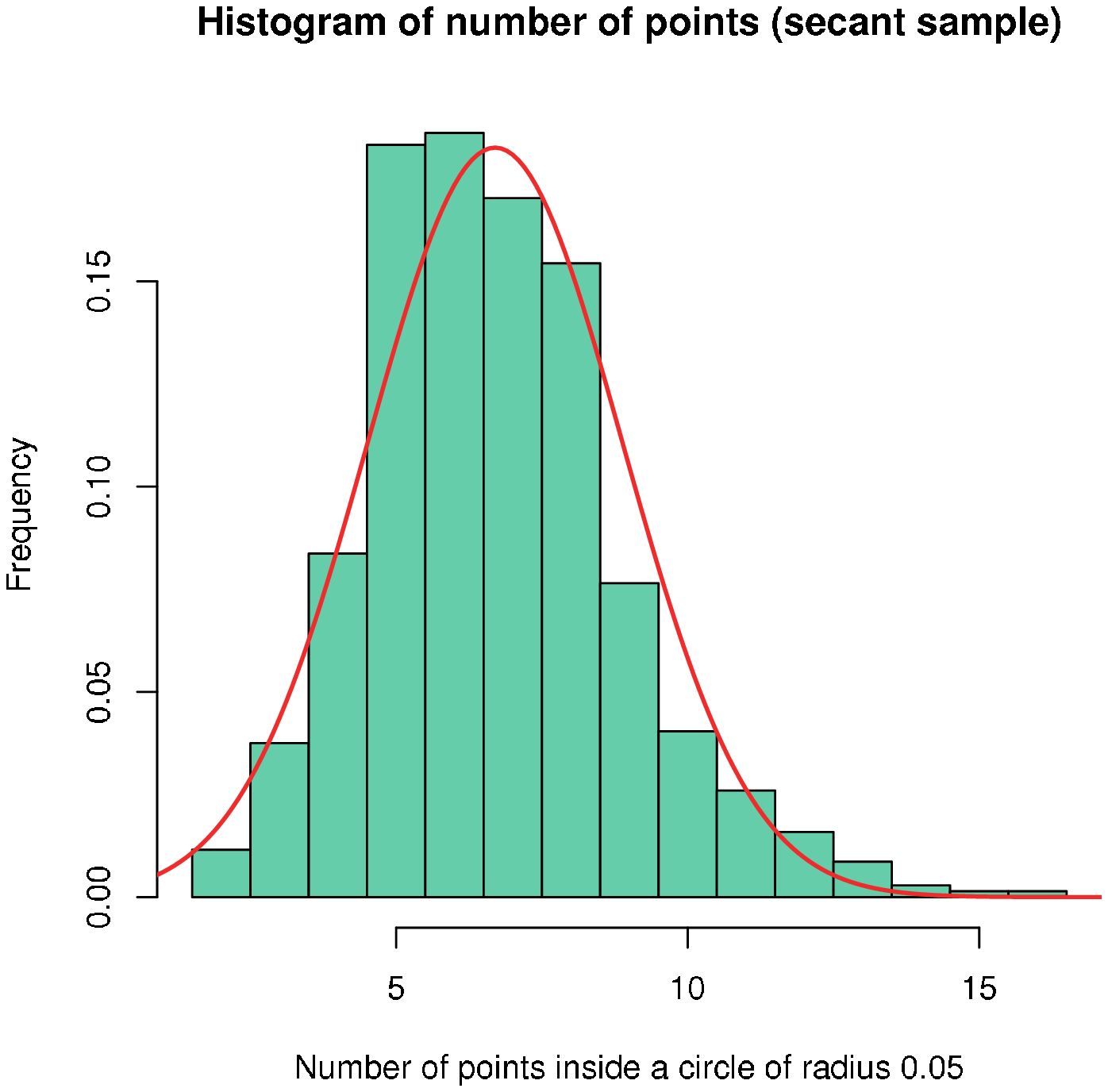}\qquad%
  \image{7}{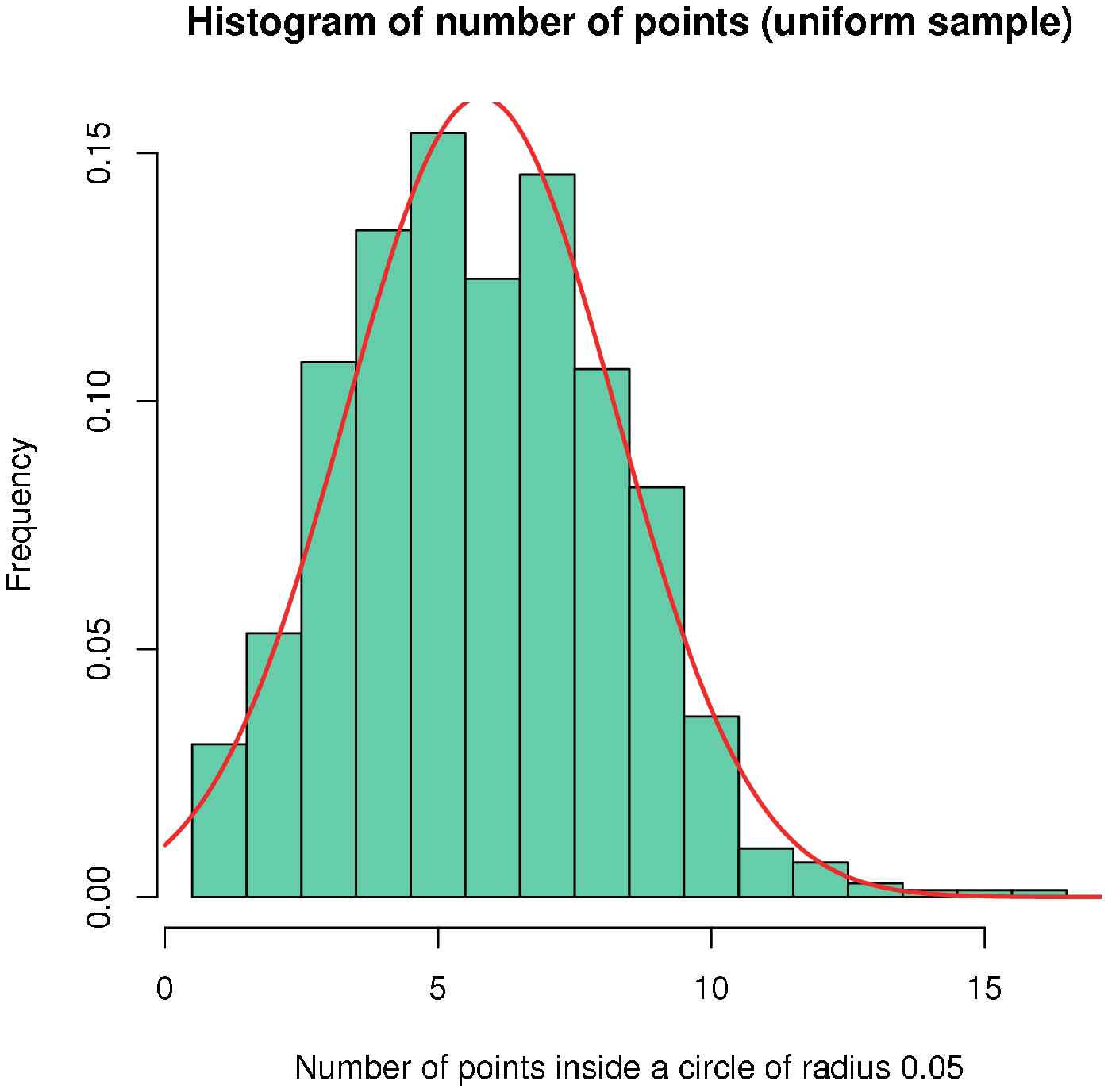}%
  \caption{Histograms of the number of points inside a random circle
    in the geometric (left) and simulated (right) cases.  Both are
    well described by a Gaussian; for a comparison, see text.}
  \label{fig:histograms}
\end{figure}
Fig.~\ref{fig:histograms} shows the result presented as two histograms
of the number of inner points for $10^3$ circles.  These can be well
approximated by Gaussians, and we can see that the variance of the
simulated sample is bigger than the variance of the geometric one: that
is, the geometric points cover hyperplane space with a more regular
(less random) pattern, in such a way that each point tries to avoid
the clusters and tends to fill the gaps, a characteristic behavior of
the so-called quasi Monte Carlo (qMC) sequences \cite{PrimNum}.

Putting the comparison in more a quantitative fashion, the mean and
standard deviation of the geometric and simulated samples are
\begin{equation}
  \label{eq:rh-15}
  \begin{split}
    \mu_{\text{Geo}} &= 6.69697\,,\\
    \sigma_{\text{Geo}} &= 2.185198\,,
  \end{split}
  \qquad
  \begin{split}
    \mu_{\text{Sim}} &= 5.787115\,,\\
    \sigma_{\text{Sim}} &= 2.474453\,.
  \end{split}
\end{equation}
The different mean values with the same number of points indicate that
the density of points is not the same in both samples.  This happens
because the spatial correlations in the geometric case make the left
and right sides underpopulated (see fig.~\ref{fig:sec-vs-sim}), so
that the effective volume covered in this case is smaller.  We can
take this into account by considering the quotient between the
standard deviation and the mean as a measure of the clustering of the
data, which is an effective scaling of the data to have mean value 1.
This leads to
\begin{equation}
  \label{eq:rh-16}
  \frac{\sigma_{\text{Geo}}}{\mu_{\text{Geo}}} = 0.3262966\,,
  \qquad
  \frac{\sigma_{\text{Sim}}}{\mu_{\text{Sim}}} = 0.4275798\,.
\end{equation}
This means that the fluctuation in the point density is greater for
the simulated sample.  The relative difference is considerable
($\frac{0.4275798 - 0.3262966}{0.3262966} = 0.3104023$, that is,
31\%).  This indicates that the covering of the hyperplane space by
the geometric sample is much more efficient than the covering made by
the simulated one, and thus its randomness is much smaller.

Summarizing, the approximation made by assuming uniformity in the
distribution of secant hyperplanes is equivalent to neglecting the
spatial correlations present in the geometric case, and we can expect
this approximation to work well because of the more efficient covering
of hyperplane space in this case.

\section{Number of states in the Weinberg Window}
\label{sec:N_WW}

The next step in our calculation is to compute the distribution of
values of the cosmological constant.  The random hyperplane model
allows us to do this by means of the marginal probability distribution
in the $\rho$ variable, whose density will be called $\omega(\rho)$:
\begin{equation}
  \label{eq:rh-17}
  \int_{\upsilon\in S^{J-1}}\dif P(\rho,\upsilon) \equiv \omega(\rho)\dif\rho\,.
\end{equation}
Using the following relation between $\rho$ (the minimum distance from
a tangent hyperplane to their secant state), $R=\sqrt{2|\Lambda_0|}$
(the radius of the $\Lambda=0$ sphere) and the Euclidean norm of the
secant state in flux space (which in turn is related to the
cosmological constant)
\begin{equation}
  \label{eq:rh-18}
  \left.
    \begin{split}
      \displaystyle\sqrt{\sum^J_{i=1}q_i^2n_i^2} = R + \rho\\
      \Lambda = \Lambda_0 +
      \frac{1}{2}\displaystyle\sum^J_{i=1}q_i^2n_i^2
    \end{split}
  \right\}
  \quad\Rightarrow\quad
  \rho(\Lambda) = \sqrt{2(\Lambda - \Lambda_0)} - \sqrt{2|\Lambda_0|}\,,
\end{equation}
the $\Lambda$ distribution can be found once the function
$\omega(\rho)$ is computed:
\begin{equation}
  \label{eq:rh-19}
  f(\Lambda)\dif\Lambda = \omega[\rho(\Lambda)]
  \frac{\dif\rho}{\dif\Lambda} \dif\Lambda
  = \omega[\rho(\Lambda)]
  \frac{\dif\Lambda}{\sqrt{2(\Lambda - \Lambda_0)}}\,.
\end{equation}
The $\omega$ function is easily computed once the integration of the
characteristic function in the $(\rho,\upsilon)$ variables is
reversed.  The $\rho$ variable takes a maximum value
$\sigma_{\text{max}} = \frac{1}{2}\sqrt{\sum^J_{i=1}q_i^2}$, and the
$\upsilon$ variable can range across $S^{J-1}$.  So we define the
integral of an arbitrary test function $\phi(h)$ on the hyperplane
space $H_Q$ against the probability measure $\dif P(h)$ as
\begin{equation}
  \label{eq:rh-20}
  \int_{H_Q}\phi(h)\dif P(h) =
  \frac{1}{K}
  \int_{(\rho,\upsilon)\in[0,\sigma_{\text{max}}]\times S^{J-1}}
  \chi_{H_Q}(\rho,\upsilon) \phi(\rho,\upsilon)
  \dif\rho\dif\Omega_{J-1}(\upsilon)\,.
\end{equation}
For the iterated integrals, we have the following identity\footnote{In
  equation \eqref{eq:rh-21}, we drop the constant $K$ computed in
  \eqref{eq:rh-13}.} (Fubini's theorem):
\begin{equation}
  \label{eq:rh-21}
  \int_0^{\sigma_{\text{max}}}\dif\rho \int_{\mathcal{J}_\rho}
  \dif\Omega_{J-1}(\upsilon)\ {} \phi(\rho,\upsilon)
  =
  \int_{S^{J-1}}\dif\Omega_{J-1}(\upsilon) \int_{\mathcal{I}_\upsilon}
  \dif\rho\ {} \phi(\rho,\upsilon)
  \,,
\end{equation}
where the sets $\mathcal{I}_\upsilon\subset[0,\sigma_{\text{max}}]$ and
$\mathcal{J}_\rho\subset S^{J-1}$ appear owing to $\chi_{H_Q}$, and are
defined as
\begin{equation}
  \label{eq:rh-22}
  \begin{split}
    \mathcal{I}_\upsilon &= \bigl\{
    \rho\in[0,\sigma_{\text{max}}]\colon (\rho,\upsilon)\in H_Q
    \bigr\}
    = \bigl\{
    \rho\in[0,\sigma_{\text{max}}]\colon 0 \le \rho \le \sigma(\upsilon)
    \bigr\}
    = [0,\sigma(\upsilon)]\,,\\
    \mathcal{J}_\rho &= \bigl\{
    \upsilon\in S^{J-1}\colon (\rho,\upsilon)\in H_Q
    \bigr\}
    = \bigl\{
    \upsilon\in S^{J-1} \colon \sigma(\upsilon) \ge \rho
    \bigr\}
    \,.
  \end{split}
\end{equation}
Note that the restrictions in both sets are the same,
$\rho\le\sigma(\upsilon)$, but in $\mathcal{I}_\upsilon$ the direction
$\upsilon$ remains fixed, and in $\mathcal{J}_\rho$ the distance
$\rho$ remains fixed.  So, $\mathcal{I}_\upsilon$ is simply the
interval $[0,\sigma(\upsilon)]$ used in the integral \eqref{eq:rh-14},
and $\mathcal{J}_\rho$ is a set which is the whole sphere when
$\rho\le \frac{1}{2}\min\{q_j\}$ and a $2^J$-connected-component set
when $\rho>\frac{1}{2}\max\{q_j\}$.  If we carry out the $\upsilon$
integration of $\dif P(\rho,\upsilon)$ but we stop before the $\rho$
integration, we have
\begin{equation}
  \label{eq:rh-23}
  \omega(\rho) = \frac{1}{K}\int_{\mathcal{J}_\rho}
  \dif\Omega_{J-1}(\upsilon) \,.
\end{equation}
To compute the integral in \eqref{eq:rh-23} in closed form is not an
easy task, and even if it were, it would be useless, because the
function $\omega$ is built by gluing polynomial functions in a
$J$-dependent number of intervals.

\begin{figure}
  \centering
  \image{7}{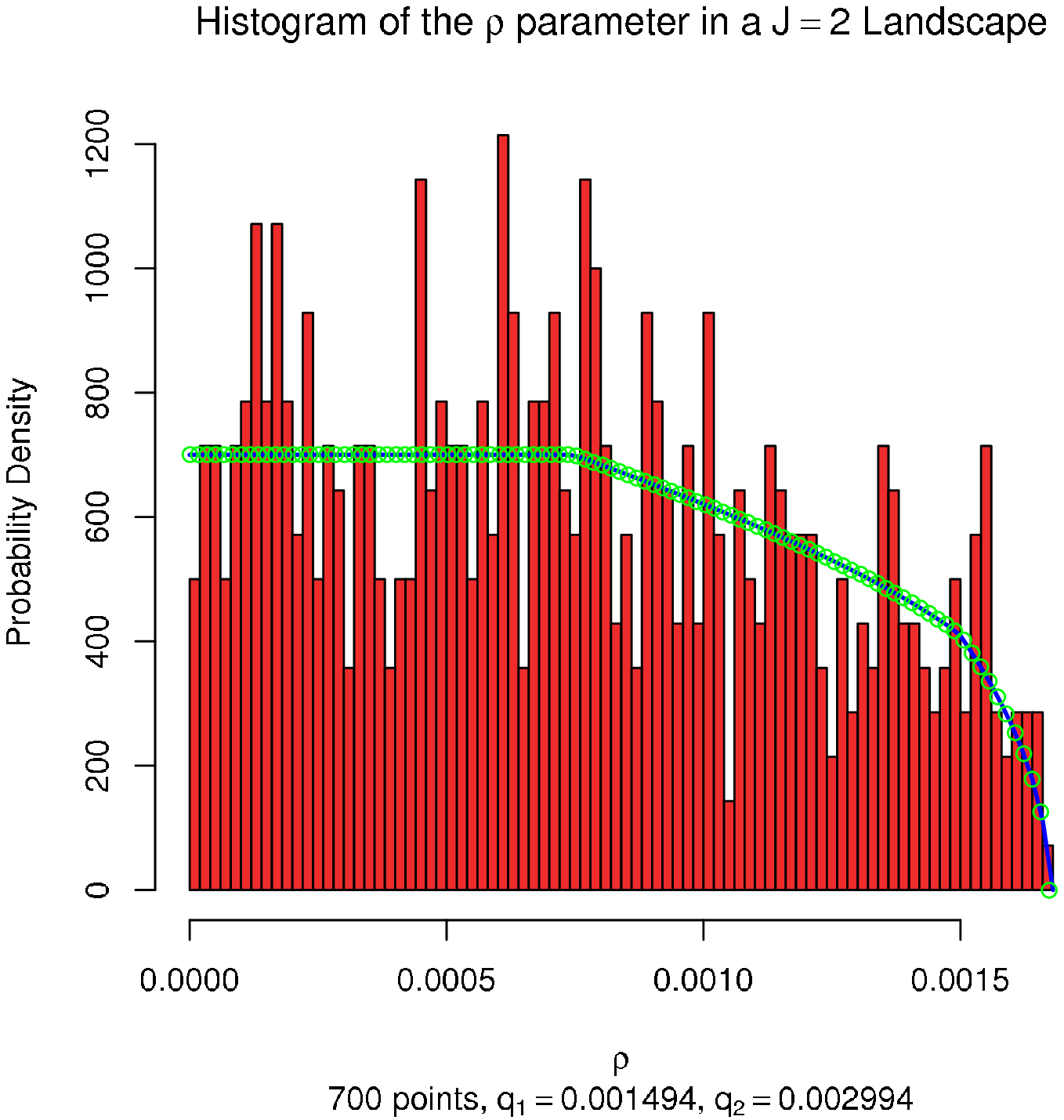}\qquad%
  \image{7}{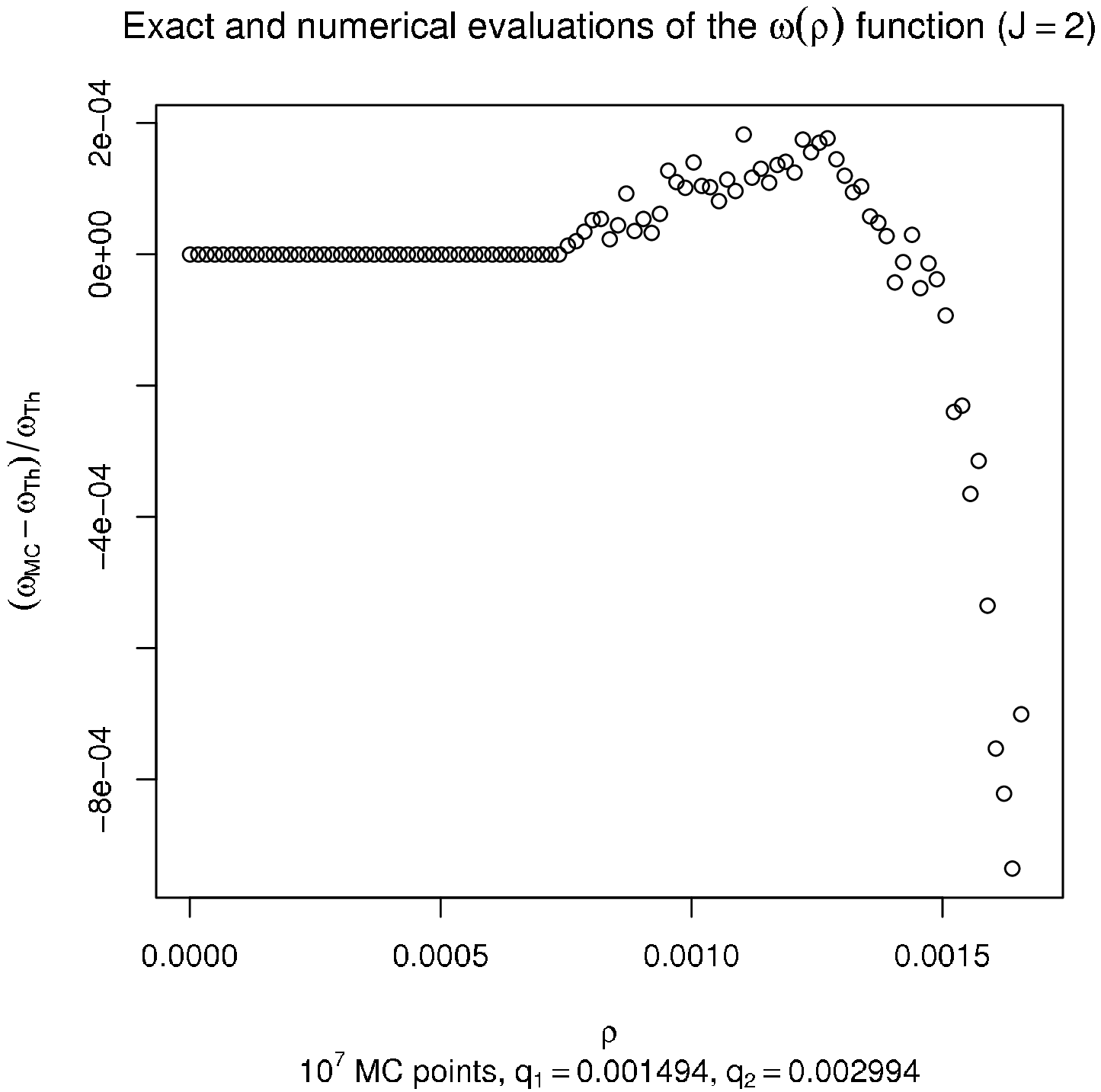}%
  \caption{Histogram of the $\rho$ parameter in a $J=2$ Landscape.
    The solid line is the $\omega$ function computed using
    eq.~\eqref{eq:rh-27}, the points represent the Monte Carlo
    computation of the $\omega$ integral (left).  Relative error
    between the numerical and exact evaluations of the $\omega$
    function in the $J=2$ case.  The MC integration used $10^7$ points
    in the circle $S^1$ (right).}
  \label{fig:omega-J2-and-MC-vs-Th}
\end{figure}

Let us compute $\omega$ for the simplest $J=2$ case.  The set
$\mathcal{J}_\rho$ is here an interval, which can be computed from the
$\sigma$ function which represents the upper boundary of hyperplane
space, see fig.~\ref{fig:sigma}.  A horizontal line at height $\rho$
has two, one or no intersection with $\sigma(\upsilon)$ in the first
quadrant depending on $\rho$.  Taking $q_1 < q_2$, there are no
intersection points if $\rho < \frac{q_1}{2}$, there is a single
intersection point if $\frac{q_1}{2} < \rho < \frac{q_2}{2}$ and there
are two intersection points if $\frac{q_2}{2} < \rho <
\sigma_{\text{max}}$.  The equation to be solved for the intersection
points in the first quadrant is (remember that in $J=2$ we have
$\upsilon = (\cos\theta,\sin\theta)$)
\begin{equation}
  \label{eq:rh-24}
  \rho = \frac{1}{2}\bigl[q_1\cos\theta + q_2\sin\theta\bigr]\,.
\end{equation}
Setting $x=\cos\theta$, we have a quadratic equation whose solution
has two branches:
\begin{equation}
  \label{eq:rh-25}
  x_\pm = \frac{1}{\sigma_{\text{max}}^2}
  \Bigl[
  \frac{q_1}{2}\rho \pm \frac{q_2}{2}
  \sqrt{\sigma_{\text{max}}^2 - \rho^2}
  \Bigr]\,.
\end{equation}
These branches only make sense for $\rho \ge \frac{q_1}{2}$ (for
$x_+$) and for $\rho \ge \frac{q_2}{2}$ (for $x_-$).  We define
$\theta_\pm = \cos^{-1} x_\pm$, so we have, in the first quadrant
only,
\begin{equation}
  \label{eq:rh-26}
  \mathcal{J}_\rho =
  \begin{cases}
    [0,\frac{\pi}{2}] & \text{if $\rho \le \frac{q_1}{2}$},\\
    [\theta_+,\frac{\pi}{2}] & \text{if $\frac{q_1}{2} \le \rho \le \frac{q_2}{2}$},\\
    [\theta_+,\theta_-] & \text{if $\frac{q_2}{2} \le \rho \le
      \frac{1}{2}\sqrt{q_1^2 + q_2^2} = \sigma_{\text{max}}$}.\\
  \end{cases}
\end{equation}
The $\omega$ function is, then,
\begin{equation}
  \label{eq:rh-27}
  \omega(\rho) = \frac{1}{K}
  \begin{cases}
    2\pi & \text{if $\rho \le \frac{q_1}{2}$},\\
    2\pi - 4\theta_+(\rho) & \text{if $\frac{q_1}{2} \le \rho \le \frac{q_2}{2}$},\\
    4[\theta_-(\rho)- \theta_+(\rho)] & \text{if $\frac{q_2}{2} \le \rho \le
      \frac{1}{2}\sqrt{q_1^2 + q_2^2} = \sigma_{\text{max}}$}.\\
  \end{cases}
\end{equation}
With $J=2$, $K = 2(q_1+q_2)$, and taking $\Lambda_0 = -1$ and charges
$q_1=0.001494$, $q_2=0.002994$, we can plot this function and compare
it with the true $\rho$ histogram of the secant states; this is done
in figure \ref{fig:omega-J2-and-MC-vs-Th} (left), where we see big
fluctuations of the number of states versus the theoretical $\omega$
function.
This strong fluctuation is a consequence of the relative sparsity of
this Landscape.  The small sample size makes the histogram oscillate
around the theoretical curve; we believe that this effect is only
present in $J=2$, as can be seen in the subsequent figures.

In figure \ref{fig:omega-J2-and-MC-vs-Th} (right) we can see the
relative error between the exact formula \eqref{eq:rh-27} and the
Monte Carlo evaluation explained below (eq.~\eqref{eq:rh-29}).  They
agree in the constant region, and the error remains smaller than
$10^{-3}$ for all values.

We can compute the values of the $\omega$ function numerically by
rewriting it in the following way, using the unit step function
$\theta(x)$ for restricting the integrand to $\mathcal{J}_\rho$ and
the formula \eqref{eq:rh-13} for the normalization constant $K$:
\begin{equation}
  \label{eq:rh-28}
  \omega(\rho) = \frac{\vol S^{J-1}}{\overline{q}\,\vol S^{J-2}}
  \int_{S^{J-1}} 
  \theta[\sigma(\upsilon)-\rho]
  \frac{\dif\Omega_{J-1}(\upsilon)}{\vol S^{J-1}}
  \,.
\end{equation}
Thus, the formula splits in a constant times the mean value of the
step function with respect to the uniform probability measure on the
sphere.  This mean value is easily computed using a simple Monte Carlo
technique, that is, sampling the unit sphere with a large number $N$
of points $\upsilon^{(i)}$, we have
\begin{equation}
  \label{eq:rh-29}
  \omega(\rho) \approx
  \frac{\vol S^{J-1}}{\overline{q}\,\vol S^{J-2}}
  \ {}\frac{1}{N}\sum^N_{i=1}\theta[\sigma(\upsilon^{(i)})-\rho]\,.
\end{equation}
The Monte Carlo evaluation is well suited for this task, because the
integrand is bounded and the integration domain is compact.

\begin{figure}
  \centering
  \image{7}{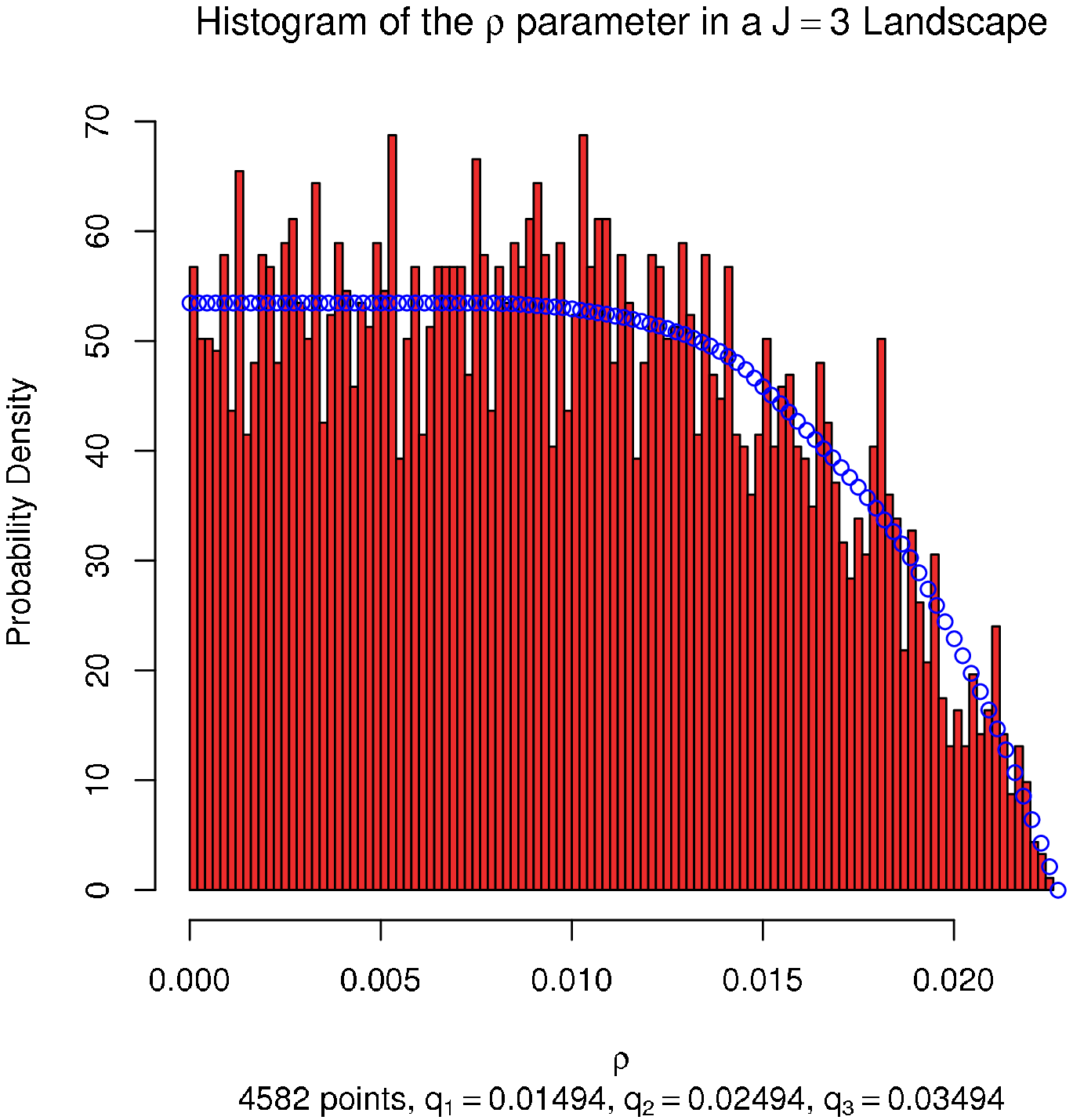}\qquad%
  \image{7}{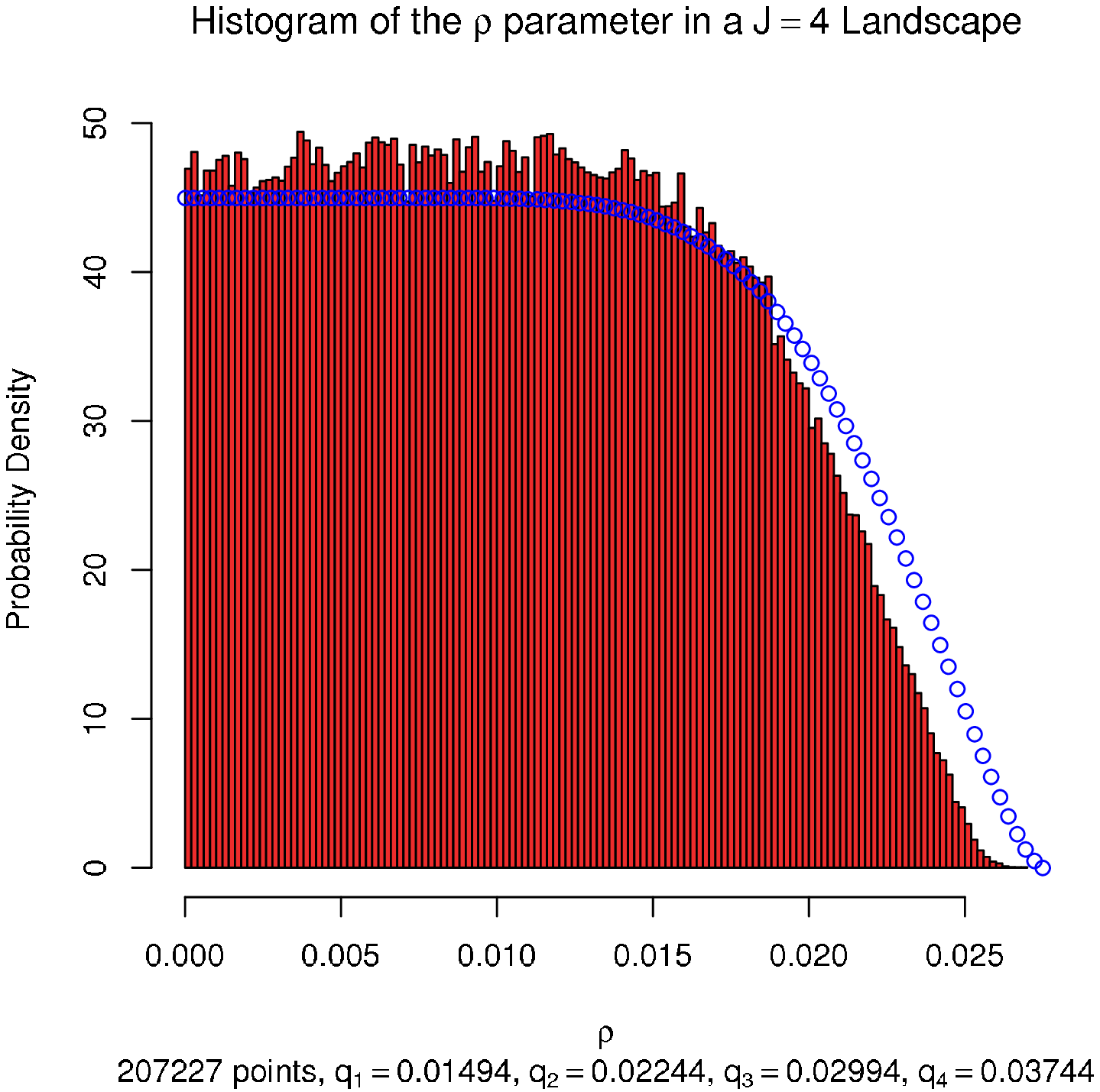}%
  \caption{Histograms of brute-force search for secant states compared
    with the MC-computed $\omega$ curve for Landscapes with $J=3$
    (left) and $J=4$ (right).  See text for an explanation of the
    differences.}
  \label{fig:omega-J3-J4}
\end{figure}

In fig.~\ref{fig:omega-J3-J4} we can see the MC-computed $\omega$
distribution for Landscapes with $J=3$ and $J=4$, where brute-force
data are available.  The $J=3$ oscillates much more than the $J=4$
one, due in part to the bin width, which contributes to make the
histogram smoother with a bigger sample size.  But the mean value in
the constant region (as well as the global fit to the entire
theoretical curve) seems to be better adjusted in the $J=3$ case.  The
reason for this apparent disagreement between the brute-force data in
$J=4$ and the theoretical MC-computed $\omega$ curve is the sampling
method used to find the secant states in this case.  This sampling
method, described above (see section \ref{sec:N_S}), has the
disadvantage of missing the states with very small area of the
intersection polytope between the $\Lambda=0$ sphere and the Voronoi
cell of the secant state.  Using eq.~\eqref{eq:c0-17}, we can estimate
the number of missing states; with $\Lambda_0=-1$, $q_1=0.01494$,
$q_2=0.02244$, $q_3=0.02994$, $q_4=0.03744$, we have
\begin{equation}
  \label{eq:rh-33}
  \mathcal{N}_{\mathcal{S},J=4}^{\text{theo}} = 6,605,383\text{ secant states.}
\end{equation}
The brute force calculation yielded 406,715 states in the first
4-quadrant; taking into account degeneracies (the degeneracy of each
state is $g=2^{4-z}$, being $z\in\{0,1,2,3\}$ the number of zero-flux
components), we obtain
\begin{equation}
  \label{eq:rh-34}
  \mathcal{N}_{\mathcal{S},J=4}^{\text{brute-force}} = 6,245,948\text{ secant states.}
\end{equation}
Their difference relative to the mean degeneracy is
\begin{equation}
  \label{eq:rh-35}
  \frac{\mathcal{N}_{\mathcal{S},J=4}^{\text{theo}}
    - \mathcal{N}_{\mathcal{S},J=4}^{\text{brute-force}}}{\langle
    g\rangle}  = 23,405.19\text{ first 4-quadrant secant states.}
\end{equation}
These states amount to 5.4\% of the total, and they lie completely in
the high-$\rho$ region.  This is the reason why the theoretical curve
disagrees with the histogram in the $J=4$ case:  the histogram is
normalized to have total area one and it misses 5\% of states of
higher $\rho$, so we must conclude that the data range of the
histogram must be shorter and its constant region must be higher, as
shown in fig.~\ref{fig:omega-J3-J4}.

But we don't need the numerical evaluation to analyze the $\omega$
function.  From eq.~\eqref{eq:rh-27} we can see that when
$\rho\le\frac{1}{2}\min\{q_i\}$, the value of the step function is
always 1, so we have the exact result
\begin{equation}
  \label{eq:rh-30}
  \omega(\rho) = 
  \frac{\vol S^{J-1}}{\overline{q}\,\vol S^{J-2}}
  =
  \frac{\sqrt{\pi}\,\Gamma\bigl(\frac{J-1}{2}\bigr)}
  {\overline{q}\,\Gamma\bigl(\frac{J}{2}\bigr)}
  \quad
  \text{for $\rho\le\frac{1}{2}\min\{q_i\}$}\,,
\end{equation}
that is, the distribution of $\rho$ values is \emph{exactly} constant
at small $\rho$, $\rho\le\frac{1}{2}\min\{q_i\}$.  Beyond this value,
the step function finds regions where $\sigma(\upsilon)<\rho$, and the
mean value begins to decrease monotonically.  When
$\rho\ge\sigma_{\text{max}}$, no point in the sphere is captured by
the step function, and $\omega$ vanishes.  We have checked this
general behavior on high-dimensional BP models before and after
performing the BT decay chain and this qualitative profile is quite
robust \cite{AS}.

As the authors of \cite{SV} point out, actual histograms of $\Lambda$
on particular BP Landscapes show a ``staggered'' behavior which gets
smoothed when the bin width increases, but the mean value of these
oscillations is given by $\omega$.  Note that $\Lambda$ is really a
\emph{discrete} variable, and $\omega$ is the density of a continuous
one.  Furthermore, note that in \cite{SV} the authors compile
statistics on entire Landscape instances, while our statistics refer
to the positive $\Lambda$ secant state sector only.

So we find that $\omega$ behaves as a Fermi-like distribution on the
$\rho$ values, with a medium filling level and a decreasing interval
which depend on the values of the charges.  Further quantitative
analysis must be done trough the numerical estimation procedure
described above.

Now, we can compute the probability for the cosmological constant to
lie in the Weinberg Window, using the properties of $\omega$ and
formula \eqref{eq:rh-19}:
\begin{equation}
  \label{eq:rh-31}
  \begin{split}
    P(0\le\Lambda\le\lww) &= \int_0^{\lww}f(\Lambda)\dif\Lambda =
    \frac{\vol S^{J-1}}{\overline{q}\,\vol S^{J-2}}
    \underbrace{
      \int_0^{\lww}\frac{\dif\Lambda}{\sqrt{2(\Lambda-\Lambda_0)}}
    }_{\rho(\lww) \approx \frac{\lww}{\sqrt{2|\Lambda_0|}} = \frac{\lww}{R}} \\
    &=
    \frac{\vol S^{J-1}\lww}{R\,\overline{q}\,\vol S^{J-2}}
    \,.
  \end{split}
\end{equation}
After inserting this in formula \eqref{eq:rh-1}, together with
eq.~\eqref{eq:c0-17}, we have
\begin{equation}
  \label{eq:rh-32}
  \num{WW}
  = \frac{1}{2}\times
  \frac{2 R^{J-1}\overline{q} \vol S^{J-2}}{\vol Q}\times
  \frac{\vol S^{J-1}\lww}{R\,\overline{q}\,\vol S^{J-2}}
  = 
  \frac{R^{J-2}\vol S^{J-1}\lww}{\vol Q}
  \,.
\end{equation}
This result exactly coincides with the BP count (eq.~\eqref{eq:rh-3}).

\begin{figure}
  \centering
  \image{9}{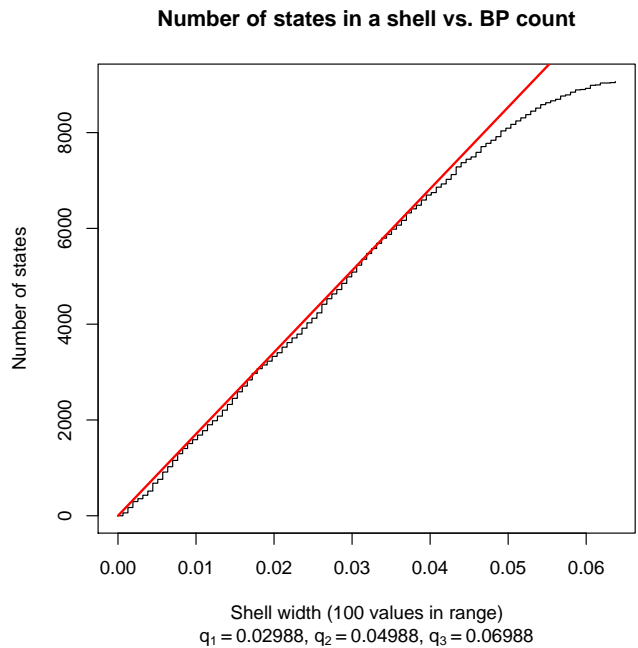}
  \caption{Comparison between an actual count of the number of states
    in a shell of variable width (stair-like line) and the number
    predicted by the BP count (plain line).}
  \label{fig:comp-stair-line}
\end{figure}

One may wonder if the approximations made by the random hyperplane
model are too crude to achieve a better result than the BP count, or
if the BP count is a better result than expected.  In order to clarify
this point, we have again performed a numerical experiment to compare
the number of states predicted by the BP (and our) count with a
brute-force search of states in a shell of variable width (that is,
varying $\lww$) in a $J=3$ Landscape with 2,333 secant states in the
first octant.  In our opinion, the nature of the example chosen would
render the approximations made too crude, but we find a remarkable
agreement, as shown in fig.~\ref{fig:comp-stair-line}, and in a range
of shell widths much wider than expected.

\begin{figure}
  \centering
  \image{7}{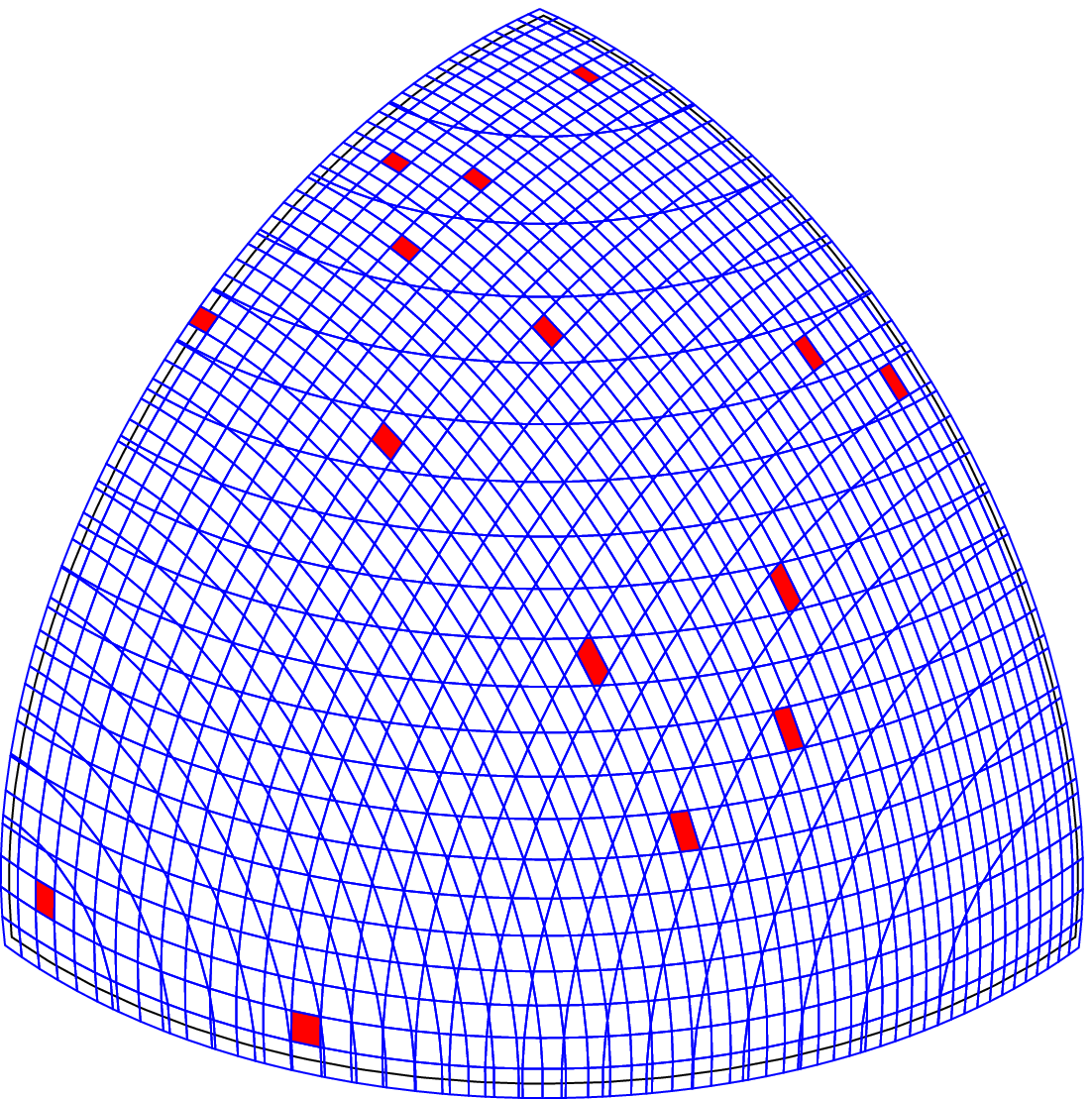}\quad%
  \image{7}{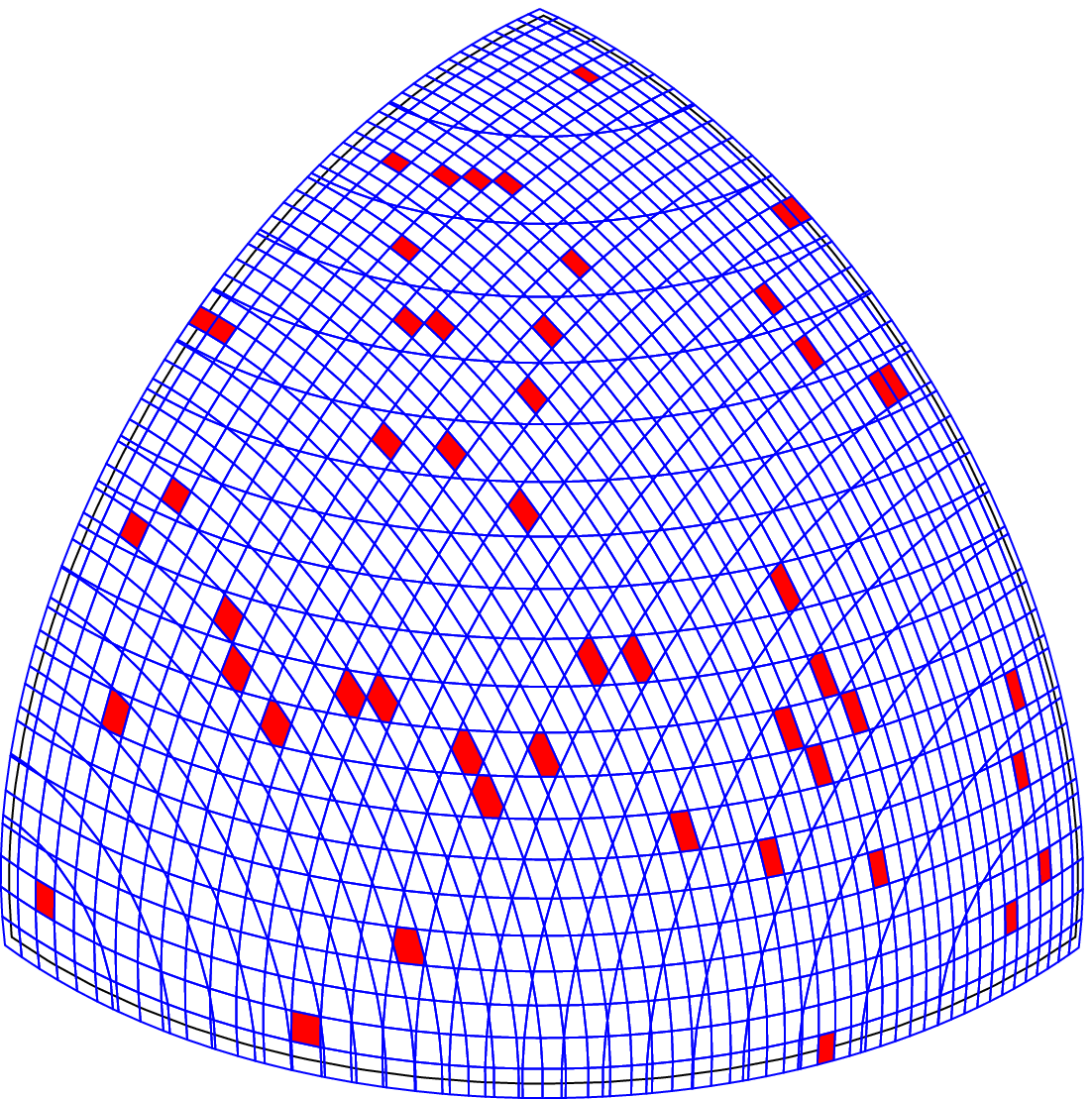}\vspace{4mm}
  \image{7}{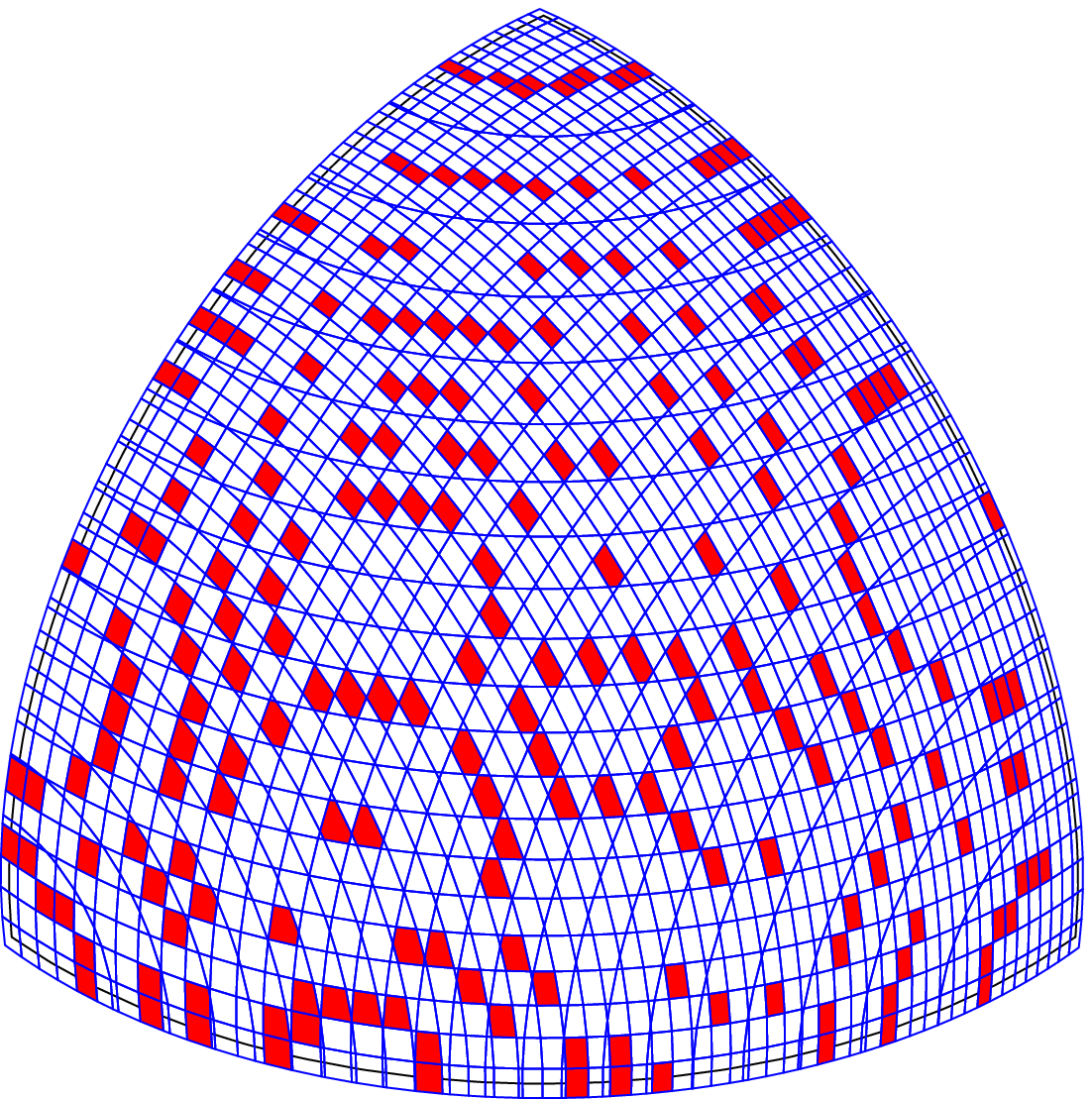}\quad%
  \image{7}{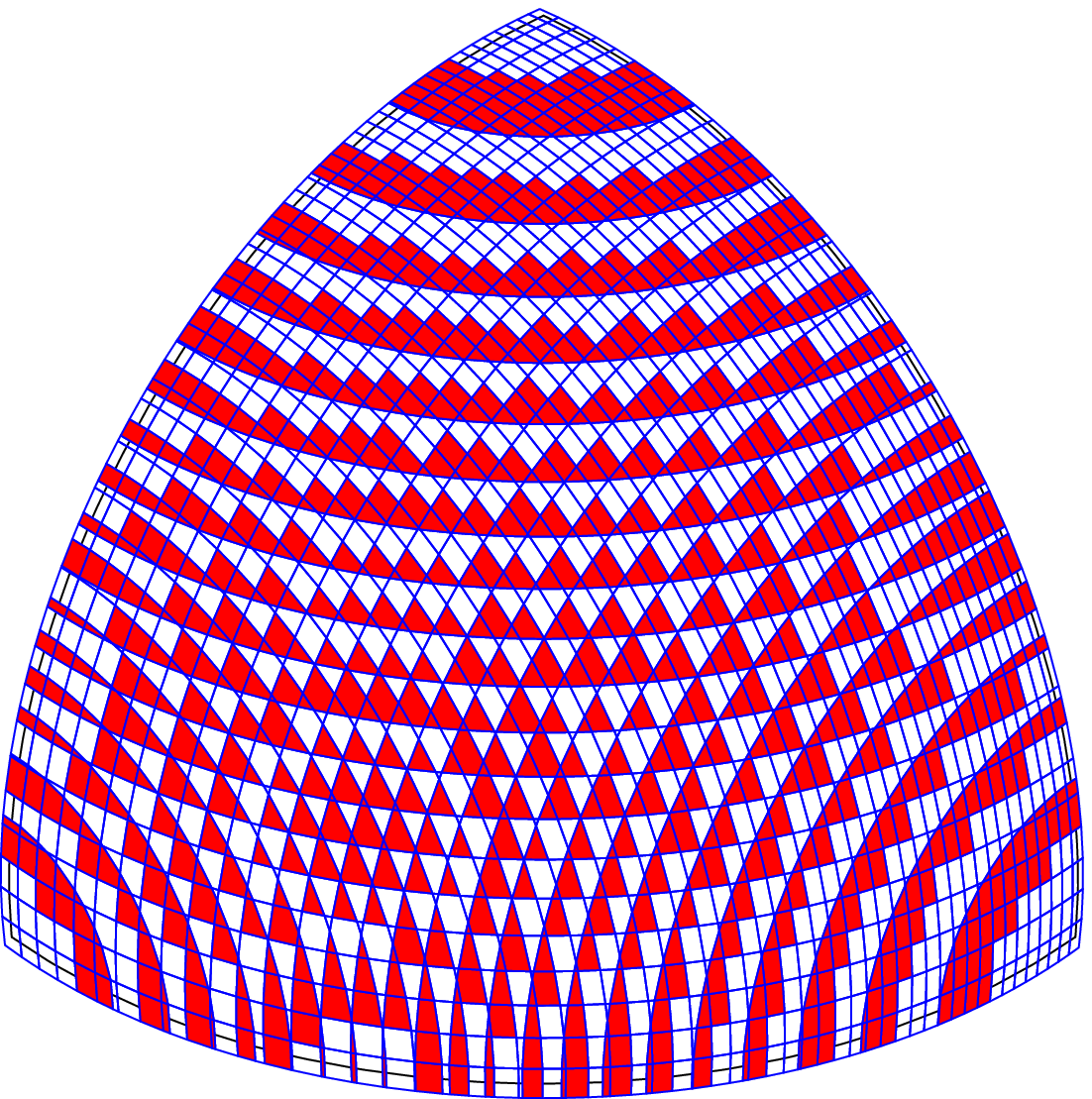}
  \caption{Blue lines represent the tessellation induced by the
    Voronoi cells of the secant states in the first octant (whose
    boundary is depicted by a black line) of the $\Lambda=0$ sphere.
    Each tessella corresponds to a secant state.  The red tessellas
    have centers inside a shell of width 0.001 (top left), 0.003 (top
    right), 0.01 (bottom left) and 0.07 ---all secant states with
    $\Lambda>0$--- (bottom right).}
  \label{fig:tessellas}
\end{figure}

We are forced to conclude that the BP count has proved to be a
succesful way of counting states in the BP Landscape, in fact better
than expected.  The geometric interpretation is curious; if we sum up
the volume of the Voronoi cells of the secant states whose center are
inside a shell of prescribed width, we obtain the volume of the shell
with very good approximation.  In figure \ref{fig:tessellas} we
illustrate the spatial location of the states inside the shell for
chosen shell widths in the $J=3$ Landscape of the previous example.
These locations are uncorrelated for low values of the shell width.

\section{Conclusions and future directions}
\label{sec:conc}

We have modeled the BP Landscape by means of the random hyperplane
model (RHM), validated by several numerical experiments like
brute-force search and precise state counting.  The RHM provides us
with a distribution of values of the cosmological constant and
reproduces the BP count in a very different way.  The numerical
experiments performed suggest that the BP count is much more precise
than expected, far better than the BY count, for example, and
applicable to a wide range of parameters of the Landscape.

Further corrections to this formula can take into account the
asymmetry of the $\Lambda>0$ and $\Lambda<0$ states due to the
curvature of the $\Lambda=0$ sphere (neglected here), which is a
second order effect which depends on the precise values for the
charges.  For us, it is certainly more difficult to take into account
the spatial correlations, which would require a completely different
approach.

All these considerations lie in the context of finding an \emph{a
  priori} distribution of the cosmological constant, without any
reference to the dynamics of the BT decay chain.  In our future
research we will try to incorporate the dynamical effects as well, in
order to count also the states selected by dynamics and to quantify
the dynamical selection effect within this framework.  In this
respect, we have checked by extensive numerical simulations that the
Fermi-like profile of the $\omega$ ditribution on the secant states
persists even on a subset selected by the BT decay chain \cite{AS}.
If this effect reveal a feature of the dynamical selection or is
caused by the geometric structure of the set of secant states is a
point which is worth of further investigation.

\appendix

\section{Improving the BP count}
\label{sec:imp}

We can slightly improve the result obtained using eq.~(\ref{eq:rh-1})
and the random hyperplane model (RHM) described above.  The first
heuristic approach is to replace the $\frac{1}{2}$ factor in
(\ref{eq:rh-1}) by a measure of the volume of a shell above and below
the $\Lambda=0$ surface in flux space.  This surface is a sphere of
radius $R=\sqrt{2|\Lambda_0|}$.  We will call the inside region of
this sphere $\mathcal{B}^J(R)$ and its volume $\vol\mathcal{B}^J(R) =
V^J(R) = \frac{R^J}{J}\vol S^{J-1}$.  We can draw a shell of width
$\epsilon$ above the sphere and another shell of the same width below.
Then, the relative volume $\eta$ of the positive-$\Lambda$ region in
the shell is given by the quotient of the volume of the upper shell
and the total volume of the two shells:
\begin{equation}
  \label{eq:rh-36}
  \eta = \frac{V^J(R+\epsilon)-V^J(R)}
  {V^J(R+\epsilon)-V^J(R-\epsilon)}
  = \frac{\bigl(1+\frac{\epsilon}{R}\bigr)^J - 1}
  {\bigl(1+\frac{\epsilon}{R}\bigr)^J - \bigl(1-\frac{\epsilon}{R}\bigr)^J}
  \,.
\end{equation}
In the limit $\epsilon\to0$ or $R\to\infty$, $\eta\to\frac{1}{2}$
which is eq.~(\ref{eq:rh-1}), and for the special case $\epsilon=R$
(when the inner shell completely fills $\mathcal{B}^J(R)$)
$\eta=1-2^{-J}$, which is arbitrarily close to 1 for high enough $J$.
So we are replacing the $\frac{1}{2}$ by a quantity $\frac{1}{2} <
\eta < 1$, and therefore the improvement is small.

It remains to determine what the relevant width $\epsilon$ is.  To
this end we use the RHM: the distribution $\omega(\rho)$ has a
Fermi-like profile, and thus we can define a kind of ``Fermi level''
as the width $\rho_0$ of a step function distribution that has the
same height $\omega_0 = \omega(0)$ as $\omega$ and is also normalized
to one, that is, $\omega_0\rho_0 = 1$.  Thus
$\epsilon=\rho_0=\frac{1}{\omega_0}$ can play the role of the
effective width of the secant state set.

Another approach which also takes into account the difference in the
volumes of the inner and outer shells is to modify the probability
measure in hyperplane space (\ref{eq:rh-8}) and so the RHM itself.
Note that the $\rho$ variable is the distance between the tangent
hyperplane and the secant state, so that $R+\rho$ is the radial
spherical coordinate in flux space of the point in the hyperplane
closest to the secant state.  Therefore the natural modification in
the probability measure (\ref{eq:rh-8}) would be
\begin{equation}
  \label{eq:rh-37}
  \dif P(h) = \dif P(\rho,\upsilon)
  = K^{-1}\chi_{H_Q}(h)(R+\rho)^{J-1}\dif\rho\dif\Omega_{J-1}(\upsilon)\,.
\end{equation}
In this alternative viewpoint of the RHM the $\rho$ variable can be
positive or negative, but the weights of the two possibilities are
different because the marginal distribution in the $\rho$ variable is
not symmetric:
\begin{equation}
  \label{eq:rh-38}
  \int_{\upsilon\in S^{J-1}}\dif P(\rho,\upsilon) =
  K^{-1}(R+\rho)^{J-1}\omega(\rho)\dif\rho\,,
\end{equation}
where $\omega$ is the same function as before extended to negative
values of $\rho$ by symmetry $\omega(-\rho)=\omega(\rho)$.  Now, we
change eq.~(\ref{eq:rh-1}) by
\begin{equation}
  \label{eq:rh-39}
  \num{WW} = \num{S} P(0<\Lambda<\lww)\,,
\end{equation}
because the new probability distribution carries the difference
between the positive and negative $\Lambda$ states.

As long as $\rho\ll R$, this modification will be harmless and the new
probability will be exactly $\frac{1}{2}$ the old one, recovering
(\ref{eq:rh-1}).  But if $\rho$ takes on high values, then the
asymmetry will be enormous:  the distribution (\ref{eq:rh-38}) will
develop a sharp peak in the positive $\rho$ region, far away from the
Weinberg Window.

To see how this phenomenon can happen, we can remember that $\rho$ is
bounded by its maximum value $\sigma_{\text{max}} =
\frac{1}{2}\sqrt{\sum_i q_i^2}$, which behaves as
$\frac{1}{2}\sqrt{J}\,\widetilde{q}$, being $\widetilde{q}$ the square
root of the second moment of the charges.  The extreme case
$\sigma_{\text{max}}\sim R$ can be written as
\begin{equation}
  \label{eq:rh-40}
  J\frac{\widetilde{q}^2}{|\Lambda_0|} \sim (2\sqrt{2})^2 = 8\,.
\end{equation}
But then we are leaving the condition of validity (\ref{eq:c0-19-3}),
where we have replaced $\mu$ by $\widetilde{q}$.

Note that unfamiliar things happen when conditions like
(\ref{eq:c0-19-3}) are violated.  We can see this by choosing a fixed
small typical charge $\widetilde{q}\ll R$ and increasing the dimension
$J$.  The distance from the corner of the cell and its center,
$\frac{1}{2}\widetilde{q}\sqrt{J}$, can thus reach the radius $R$,
that is, the Voronoi cell around the origin in flux space will
eventually find its corners touching the surface of the
sphere\footnote{The cell acquires a kind of ``dendritic'' structure,
  with its $2J$ faces lying far away from the sphere and its $2^J$
  corners touching it.  An enormous fraction of the volume of such a
  small, $J$-dimensional cell is located at its corners.} of radius
$R$.  At this point all states inside the $\Lambda=0$ sphere will be
secant; of course many states outside the sphere will be secant also,
and a huge fraction of them will be in the corners of a parallelotope
enclosing the sphere.  But the vast majority of these states will not
have neighbors of negative cosmological constant, and therefore the
typical value of $\Lambda$ in these states will be large, hence the
peak in the $\rho$ distribution mentioned above.

We see that once the validity condition is violated, the majority of
the secant states are no more boundary states, and thus the secant
states in this regime have no special relation with the Weinberg
Window.  But the secant state set $\mathcal{S}$ is simply a tool to
study the boundary state set $\mathcal{P}$, which comprises the states
selected by dynamics, and we can use $\mathcal{S}$ instead of
$\mathcal{P}$ only when they significantly overlap.  We believe that
this is the case when the validity conditions are satisfied.

\section*{Acknowledgments}
\label{ack}

We would like to thank Roberto Emparan for useful discussions.  We
also thank Concha Orna for carefully reading this manuscript.  This
work has been supported by CICYT (grant FPA-2006-02315) and DGIID-DGA
(grant 2007-E24/2). We thank also the support by grant A9335/07
(F\'{\i}sica de alta energ\'{\i}a: Part\'{\i}culas, cuerdas y
cosmolog\'{\i}a).

\end{document}